\documentclass[twocolumn]{aastex631}
\makeatletter
\renewcommand{\frontmatter@title@above}{}
\makeatother

\usepackage{graphicx}
\usepackage{amssymb}
\usepackage{amsmath}
\usepackage{graphicx}
\usepackage{amstext}
\usepackage{leftidx}
\usepackage{esint}
\usepackage[utf8]{inputenc}
\usepackage{color}
\usepackage{hyperref}
\usepackage{multido}

\newcommand{\rL}{\rho_\Lambda}

\newcommand{\CC}{\Lambda}

\newcommand{\rv}{\rho_{\rm vac}}
\newcommand{\Pv}{P_{\rm vac}}
\newcommand{\rvo}{\rho^0_{\rm vac}}

\newcommand{\Omo}{\Omega^0_{m}}

\newcommand{\OMo}{\Omega_{m}^0}
\newcommand{\ORo}{\Omega_{r}^0}
\newcommand{\OL}{\Omega_{\rm vac}}

\newcommand{\OLo}{\Omega^0_{\rm vac}}
\newcommand{\OX}{\Omega_{X}}
\newcommand{\OXo}{\Omega_{X}^0}

\newcommand{\OD}{\Omega_{D}}
\newcommand{\ODo}{\Omega_{D}^0}

\newcommand{\rco}{\rho^0_{c}}

\newcommand{\rD}{\rho_D}

\newcommand{\rX}{\rho_X}
\newcommand{\pX}{P_X}
\newcommand{\wX}{w_X}

\newcommand{\wm}{w_m}
\newcommand{\wR}{w_r}

\newcommand{\pD}{p_D}

\newcommand{\zm}{z_{\rm max}}

\newcommand{\weff}{w_{\rm eff}}

\newcommand{\mpl}{m_{\rm Pl}}

\newcommand{\be}{\begin{equation}}
\newcommand{\ee}{\end{equation}}
\newcommand{\jtext}[1]{{\textcolor{black}{#1}}}

\begin{document}

\title{$\Lambda$XCDM: a running vacuum strategy for crossing the  phantom divide}

\author[0000-0002-5295-8275]{Joan Sol\`a Peracaula}
\affiliation{Departament de F\'isica Qu\`antica i Astrof\'isica and Institute of Cosmos Sciences,\\ Universitat de Barcelona,
c. Mart\'{\i} i Franqu\`es 1, E-08028 Barcelona, Catalonia, Spain}

\author[0000-0002-2922-2622]{Adri\`a G\'omez-Valent}
\affiliation{Departament de F\'isica Qu\`antica i Astrof\'isica and Institute of Cosmos Sciences,\\ Universitat de Barcelona,
c. Mart\'{\i} i Franqu\`es 1, E-08028 Barcelona, Catalonia, Spain}

\author[0009-0000-8375-8172]{Alex Gonz\'alez-Fuentes}
\affiliation{Departament de F\'isica Qu\`antica i Astrof\'isica and Institute of Cosmos Sciences,\\ Universitat de Barcelona,
c. Mart\'{\i} i Franqu\`es 1, E-08028 Barcelona, Catalonia, Spain}


\begin{abstract}
Composite dynamical dark energy (DDE) has recently been explored as an efficient way to help cure cosmological tensions through the so-called $w$XCDM model\,\citep{Gomez-Valent:2024tdb,Gomez-Valent:2024ejh}, a toy-model version of the $\Lambda$XCDM model\,\citep{Grande:2006nn}. The latter is a composite running vacuum model (RVM) that involves a DE component $X$ (`cosmon') of generic nature.  We compute the effective equation of state of $\Lambda$XCDM and use state-of-the-art techniques to fit this model to two standard sets of cosmological data, one involving SNIa from Pantheon$+$ and the other SNIa from DES-Dovekie, in addition to  BAO data from DESI DR2 and the CMB data from Planck PR4. We do not use large scale structure formation data for this analysis nor the SH0ES calibration of $H_0$.
We find that $\Lambda$XCDM naturally performs the crossing of the phantom divide as observed by DESI near $z\simeq 0.4$  using the $w_0w_a$CDM parameterization, a feature well favored by existing model-agnostic analyzes of the same data \citep{Gonzalez-Fuentes:2025lei,Gonzalez-Fuentes:2026rgu}. It turns out that the cosmon $X$  behaves as `phantom matter' (PM) near the present, which  in contrast to usual phantom DE  satisfies the strong energy condition (as ordinary matter) and furnishes positive pressure ($P_X>-\rho_X>0$) at the expense of negative energy density ($\rX<0$). $\CC$XCDM provides a  better fit than $w_0w_a$CDM and, as a bonus, alleviates the cosmic coincidence problem.  Given that  PM appears in stringy versions of the RVM\,\citep{Mavromatos:2020kzj,Mavromatos:2021urx}, the $\Lambda$XCDM appears to be a composite DDE model with a good chance of explaining the crossing of the phantom divide from first principles, therefore providing theoretical support to the DESI observations inferred from generic parameterizations of the DE.
\end{abstract}




\section{Introduction} \label{sec: intro}
For a long time, the standard or concordance model of cosmology, aka $\CC$CDM, which is characterized by a rigid cosmological constant (CC) $\CC$,  has been rather successful since it has provided a fairly reasonable description of the general cosmological data \, \citep{Planck:2015fie,Planck:2018vyg}. Nevertheless, the theoretical meaning of $\CC$ has never been clarified on fundamental grounds.  This has led to a big enigma:  the ``cosmological constant problem'' (CCP), see e.g. \cite{Weinberg:1988cp,Peebles:2002gy,Padmanabhan:2002ji,SolaPeracaula:2022hpd,Sola:2013gha}  and references therein. It stems from the fact that the vacuum energy density (VED), $\rv$,  which is related to the measured $\CC$ through $\rv=\CC/(8\pi G_N)$ ($G_N$ being Newton's gravitational coupling)  is many orders of magnitude smaller than the theoretical prediction for $\rv$ in quantum field theory (QFT). Although this is deeply upsetting, it turns out that all proposed forms of dark energy (DE) replacing $\rv$ and trying to circumvent this difficulty suffer from the same kind of theoretical illness. Thus, the vacuum energy is not to be signaled as the only culprit of the CCP, since any form of DE suffers from the same problem -- a fact not sufficiently recognized in the literature\,\citep{SolaPeracaula:2022hpd}. However, it is rewarding to see that recent theoretical developments on the renormalization of the vacuum energy in QFT offer new vistas for alleviating this longstanding conundrum of theoretical physics and cosmology. In this sense, there is hope to reinstate the fundamental notion of quantum vacuum as a perfectly suitable concept to describe the observed properties of the DE on fundamental grounds, in particular its dynamical character.

The mentioned developments are related to the running vacuum model (RVM) \citep{SolaPeracaula:2026pgi,SolaPeracaula:2022hpd,Sola:2013gha} formulated within the so-called off-shell adiabatic renormalization of the VED\,\citep{Moreno-Pulido:2020anb,Moreno-Pulido:2022phq,Moreno-Pulido:2023ryo}, which implies that the quantum vacuum (and hence  $\CC$ itself) is actually a  dynamical quantity (i.e. evolving with the cosmological expansion) rather than being stuck at a rigid value. One can show that this approach can ease the CCP and can offer a unified QFT \jtext{formulation of }  DE and inflation\,\citep{SolaPeracaula:2026trz,SolaPeracaula:2025yco}. For a stringy version of the RVM with similar properties, see \cite{Mavromatos:2020kzj,Mavromatos:2021urx}, and for phenomenological applications, cf.\,\cite{Gomez-Valent:2023hov}.


Although we still ignore the ultimate nature of the DE, the data releases of the Dark
Energy Spectroscopic Instrument (DESI) indeed suggest tantalizing evidence that the DE might be a dynamical quantity \citep{DESI:2024mwx,DESI:2025zgx}, a feature that has also been pointed out in more model-independent analyzes, see, e.g., \cite{DESI:2024aqx,Jiang:2024xnu,DESI:2025fii,Berti:2025phi,Li:2025ops,Gonzalez-Fuentes:2025lei,Gonzalez-Fuentes:2026rgu}. If confirmed, this would represent a profound paradigm change with respect to the $\CC$CDM model, in which $\CC$ remains constant throughout cosmic history.  The current level of evidence of dynamical dark energy (DDE)  achieved by the mentioned analyzes reaches $2.4-3.2\sigma$, depending on the Type Ia supernovae (SNIa) compilation and dataset combination employed.  It is worth noticing that anticipatory hints of DDE at a similar level of evidence appeared more than a decade ago using the RVM framework to fit a rich panoply of SNIa+BAO+$H(z)$+LSS+CMB observations\,, see e.g.\,\cite{Sola:2015wwa,Sola:2016jky,Sola:2017znb,SolaPeracaula:2016qlq,SolaPeracaula:2017esw,Gomez-Valent:2018nib}. Subsequent studies also supported the possibility of DDE using a variety of alternative methods\,\citep{Zhao:2017cud,SolaPeracaula:2018wwm}. For more recent RVM analyzes, see \cite{SolaPeracaula:2023swx,SolaPeracaula:2021gxi}.

The impact of DDE on alleviating the cosmological tensions has also been widely scrutinized. Let us recall that among these tensions the most severe is the Hubble parameter tension, i.e. the mismatch between the value of $H_0\equiv 100 h$ km/s/Mpc obtained from  CMB measurements against that collected from distance ladder observations in the local universe.  In fact, this tension was spotted as of the first measurements of the Planck collaborations and appeared as an ever-increasing discrepancy between early universe observations and late-universe determinations of $H_0$. In its most acute episode, it is found to peak at a level of $\sim 5\sigma$ between local direct measurements, e.g. $H_0 = 73.04\pm 1.04$ kms/s/Mpc
(SH0ES,\,\cite{Riess:2021jrx}), and the global fitting constraint from CMB: $H_0 = 67.27\pm 0.60$  kms/s/Mpc (Planck 2018,\,\cite{Planck:2018vyg}). Another kind of tension appears in the growth of large scale structures (LSS) and involves the measurement of the parameters $S_8$ and $\sigma_8$, or, alternatively, the more anchored parameter $\sigma_{12}$ associated with the amplitude of the matter power spectrum at fixed spheres of radius $12$ Mpc rather than at the floating scale $8h^{-1}$ Mpc used in the definition of $\sigma_8$, which hinges on the value of $h$\,\citep{Sanchez:2020vvb,eBOSS:2021poy,Semenaite:2022unt,Forconi:2025cwp}. Despite that the recent KiDS-Legacy
cosmic-shear analysis\,\citep{Wright:2025xka} is essentially in agreement (to within $0.73\sigma$) with Planck-CMB and BOSS CMASS measurements\,\citep{Xu:2024cix}, the latest DES Y6 cosmic shear results still find a remnant tension of $2.0-2.3\sigma$ in the $S_8$ measurement \citep{DES:2026mkc}.

To alleviate the above tensions, and others, a large variety of strategies have been concocted, see e.g. \cite{Heisenberg:2022gqk,Marra:2021fvf,Alestas:2020zol,Perivolaropoulos:2021bds,Alestas:2021luu,Perivolaropoulos:2022khd,Gomez-Valent:2023uof}. The reader can find a summarized account of these tensions in\,\cite{DiValentino:2020zio,DiValentino:2020vvd}. For detailed reviews, see e.g. \,\cite{Perivolaropoulos:2021jda,Abdalla:2022yfr,Cai:2026swf,Vagnozzi:2023nrq,Vagnozzi:2019ezj} and references therein. Among the manyfold strategies entertained to mitigate these inconsistencies, the possibility of having a sign flip of the DE term at some redshift in the relatively recent past has been contemplated. This  is particularly efficient when only transversal/angular BAO (BAO 2D for short) is used in the fitting analysis. For example, the $w$XCDM model presented in \cite{Gomez-Valent:2024tdb,Gomez-Valent:2024ejh} offers an efficient perspective to ease the tensions along these lines. Such a composite model is based on the interplay between two DE components $(X,Y)$ with respective equations of state (EoS's) $w_X$ and $w_Y$, in which $X$ acts first in the past cosmic history and behaves as `phantom matter' (PM) \citep{Grande:2006nn}, namely as an effective fluid with EoS $w_X<-1$ and negative energy density ($\rho_X<0$), whereas $Y$ acts more recently and behaves like normal quintessence (hence with $w_Y\gtrsim-1$, $\rho_Y>0$).  A particular case of the $w$XCDM model is when $w_X=w_Y=-1$ and then a sudden transition from anti-de Sitter into de Sitter (dS) space occurs\,\citep{Akarsu:2021fol,Akarsu:2023mfb,Akarsu:2019hmw}), see also \cite{Bouhmadi-Lopez:2026vyc}. Unfortunately,  \jtext{the $H_0$ tension does not decrease} when anisotropic (3D) BAO is used instead of transversal one, as might be expected given the existing tension between the two BAO data sets \citep{Favale:2024sdq}. Notwithstanding this fact, it is remarkable that the global $w$XCDM fit remains substantially better than in the $\CC$CDM, even with BAO 3D\,\citep{Gomez-Valent:2024tdb,Gomez-Valent:2024ejh}. This gives a hint that composite DDE may play a significant role in better describing the cosmological data.  In turn, this observation suggests that it might also help better account for another characteristic feature of DESI observations, to wit: the crossing of the phantom divide from an effective phantom DE behavior in the past into quintessence-like behavior near our time \,\citep{DESI:2025zgx}.

Although $w$XCDM is well tailored to implement that kind of crossing,  it is actually a toy-model version of a more general framework that already existed long ago, called the  $\CC$XCDM model \, \citep{Grande:2006nn,Grande:2006qi,Grande:2008re}. In the latter, we do not have separated DE phases $X$ and $Y$ acting in sequence; rather, there exists a running vacuum that exchanges energy throughout the cosmic history with the $X$ component-- called `cosmon' in the $\CC$XCDM context\, \citep{Grande:2006nn}. In this kind of scenario, the EoS that we observe is neither that of the RVM nor that of $X$ but the effective EoS $\weff$ of the composite fluid made of the two. Such an effective EoS can behave as phantom DE in the past ($\weff<-1$) and as quintessence at present ($\weff\gtrsim-1$) and one can have a continuous transition  from $\CC<0$ to $\CC>0$. Thus, while the $w$XCDM can perform a discontinuous crossing of the phantom divide, its progenitor $\CC$XCDM is more powerful and can realize it in a continuous way. The detailed study of this continuous crossing feature within the $\CC$XCDM is actually one of the main purposes of this Letter. At the same time, we perform a comparison with the $w_0w_a$CDM parameterization (also called CPL\,\citep{Chevallier:2000qy,Linder:2002et} which was first used to pinpoint the crossing \jtext{property of the observed  DDE}.

The main motivation we have to revisit the $\CC$XCDM framework here is twofold. On the one hand, it crucially reunites basic ingredients for a successful description of the main observed features of DDE by DESI; \jtext{on the other hand}, it is framed in the RVM context, which in turn is based on QFT in curved spacetime and string theory. This robust RVM core replaces the ad hoc component $Y$ in the phenomenologically successful $w$XCDM model, although the cosmon component $X$ remains as an indispensable ingredient of the $\CC$XCDM.

The $\Lambda$XCDM framework studied here is genuinely different from models that consider interactions between dark matter and dark energy to achieve the effective crossing of the phantom divide \citep{Chakraborty:2024xas,Chakraborty:2025syu,Khoury:2025txd,Wang:2025znm,Li:2026xaz,deCruzPerez:2025dni,Gomez-Valent:2026ept}. The cosmon is of a generic nature and can be either a fundamental field or the result of an effective action behavior in generalized theories of gravity. For example, X might be identified with, say, a moduli or dilaton field left over as a low-energy “relic” in string theory; it could also be a pseudo-dilaton field (acquiring a small mass) as was the case in the original `cosmon model' proposed long ago in\,\cite{Peccei:1987mm} and further discussed by Weinberg in his famous review of the CCP\,\citep{Weinberg:1988cp}. But, in general, $X$ need not be a fundamental field, it may represent a  dynamical
contribution in the effective action of vacuum generated e.g. from functionals of the form ${\cal F}(R,G,S,Q,T...)$ which are common in generalized gravity theories of various sorts\, (see e.g.\,\cite{Sotiriou:2008rp,Capozziello:2011et,Harko:2011kv,Cai:2015emx,BeltranJimenez:2017tkd,Errahmani:2025vde}, and references therein), which may influence the effective DE value that is actually measured beyond the vacuum energy effects usually encoded in $\CC$. A particular realization of the $\CC$XCDM model along the above lines was given in \cite{Bauer:2010wj,Sola:2011qr}, where the role of $X$ is played by a functional ${\cal F}={\cal F}(R,G)$ of the curvature $R$ and the Gau\ss-Bonnet term $G$. This model provides a concocted mechanism to automatically adjust the VED to a small value at present no matter how big was in the past.

In the $\CC$XCDM approach, the EoS parameter $\weff$ which we measure is to be viewed as an
effective one calculated from the composite fluid made up of the running vacuum and the cosmon.  As indicated, it does not presume the existence of scalar fields (as e.g. in quintom models\,\citep{Cai:2025mas,Gomez-Valent:2025mfl,Goh:2025upc}  and the like\,, e.g. \cite{Akarsu:2026lva,Montani:2026bzk}), which hypothetically provide
direct physical support of the DE. This is not necessary. For instance, in the work\,\,\cite{Sola:2005et}
it was shown that a running $\CC$ model leads to an
effective EoS, $\pD=\weff\,\rD$, of the combined fluid ($\rD,\pD)$ which may behave as quintessence
or even as a phantom-like DE. More generally, in \,\cite{Sola:2005nh} it was proven that any model based on Einstein's equations with
a variable $\CC=\CC(t)$ and/or $G_N=G_N(t)$ leads to an effective EoS
parameter $\weff$ that can emulate a
scalar field in both  quintessence ($-1<\weff<-1/3$) or
phantom ($\weff<-1$) disguise, generally leading to
crossing of the phantom divide
$\weff=-1$. For subsequent studies, see \cite{Das:2005yj} and \cite{Basilakos:2013vya}.

 In addition to the various field theory and effective gravity models  mentioned above, there exist other interesting scenarios that support the structure of $\CC$XCDM in the current literature. A conspicuous example appears in the stringy RVM context \citep{Mavromatos:2020kzj,Mavromatos:2021urx}, where  we can have a \jtext{fundamental} component displaying the behavior of `phantom matter'(PM). Although the latter is characterized by $w_X<-1$, as with the `ordinary' phantom DE option, PM is highlighted by the unusual property that it carries positive pressure ($\pX=w_X\rX>0$)  and negative energy density ($\rX<0$). As we shall see, this feature proves essential for the effective EoS  behavior of the composite dark energy fluid, $\weff$, being responsible for the crossing of the phantom divide in just the precise fashion observed by DESI measurements \, \citep{DESI:2025zgx}, that is, going from ordinary phantom DE behavior in the past  towards quintessence-like behavior at present.

 In this Letter, we show that the $\CC$XCDM framework is phenomenologically favored. It can  explain the crossing of the phantom divide as well as alleviate the cosmic coincidence problem within one and the same region of parameter space, and at the same time provides a quality fit to basic cosmological data which is statistically comparable, if not better, than the fitting performance of the CPL parameterization.


\section{A pr\'ecis of $\CC$XCDM cosmology}\label{sec:CompositeDE}
The $\CC$XCDM is an RVM-born model, thus incorporating an affine function of the vacuum energy density (VED) that is quadratically evolving with the Hubble rate $H$\,\citep{SolaPeracaula:2022hpd,SolaPeracaula:2026pgi}. A key additional feature is that it involves the cosmon component $X$ of a very generic nature (see the Introduction) that plays a crucial role in combination with the running vacuum. The $\CC$XCDM  was originally conceived as a model capable of alleviating the cosmic coincidence problem. It exists in different versions since long ago\,\citep{Grande:2006nn,Grande:2006qi,Grande:2008re} but here we shall focus on the original formulation, in which only the vacuum and the component $X$ exchange energy, whereas matter does not exchange energy with any of the two DE components and hence remains self-conserved.

Being the $\CC$XCDM built up in the RVM framework, its VED  in the current universe evolves as follows\,\citep{SolaPeracaula:2022hpd}:
\begin{equation}\label{eq:RVM}
\rv(H) = \rvo+\frac{3\nu}{8\pi}\,(H^2-H_0^2)\,\mpl^2\,,
\end{equation}
with $\rvo$ the current VED value,  $\mpl$ the Planck mass  and  $|\nu|\ll1$ a small  parameter which is formally computable in QFT: it plays the role of the $\beta$-function coefficient of the running vacuum \citep{Moreno-Pulido:2020anb,Moreno-Pulido:2022phq,Moreno-Pulido:2023ryo}. For $\nu>0$ the VED decreases with the expansion and then the RVM (in the absence of the cosmon)  mimics quintessence, whereas for $\nu<0$ the VED increases with the expansion and the RVM behaves effectively as phantom DE\,\citep{Sola:2005et,Sola:2005nh}. However, in the presence of $X$, the quintessence-like or phantom-like behavior of $\CC$XCDM is determined by the effective EoS of the composed DE fluid, $\weff$, which we discuss below.

Solving the $\CC$XCDM \jtext{is a bit laborious but straightforward} (see\ \cite{Grande:2006nn} for details). In what follows, we summarize the basic equations of that model, which proves entirely analytical, provided we assume that $\wX$ is constant and that the vacuum EoS is the canonical one $\Pv=-\rv$. Although it is known that the EoS of vacuum receives quantum effects in the RVM\,\citep{Moreno-Pulido:2022upl}, we shall ignore them here in order to facilitate a fully analytical solution of $\CC$XCDM.

Let $\rD=\rv+\rX$ be the total DE density of this model and $\pD=\Pv+\pX$ the corresponding pressure. The effective EoS  of the combined fluid is
\begin{equation}\label{eq:EOS}
\weff=\frac{\pD}{\rD}=\frac{-\rv+\wX\,\rX}{\rv+\rX}=
-1+(1+\wX)\,\frac{\OX}{\OD}\,,
\end{equation}
where $\OD=\rD/\rco$ and $\OX=\rX/\rco$ are the dimensionless energy densities normalized with respect to the current critical density. From the above formula it is clear that even assuming that $X$ behaves as a barotropic fluid with constant
$\wX$, the mixture of $X$ and vacuum is non-barotropic because
$\rX$ and  $\rv$ exchange energy and are functions of time or redshift,
and so is the effective EoS  $\weff=\weff(z)$.

Because matter is assumed to be conserved, $\dot{\rho}_m+3H(1+\wm)\rho_m=0$, the total DE density in the $\CC$XCDM is also self-conserved: $\dot{\rho}_D+3H(1+\weff)\rho_D=0$. With the help of \eqref{eq:EOS}, the latter can be written
\begin{equation}\label{conslawDE2}
\dot{\rho}_{\rm vac}+\dot{\rho}_X+\,3 H (1+\wX)\rX=0\,.
\end{equation}
For constant $\rv$ (non-running vacuum), it trivially boils down to the self-conservation
of the cosmon component,  $\label{conslawX}
\dot{\rho}_X+\,3 H (1+\wX)\rX=0\,.$
However, for running $\rv$ as given by Eq. \eqref{eq:RVM}, the above conservation law for the total DE shows that the
dynamics of vacuum and cosmon become closely intertwined. In fact, solving for their energy densities (assuming spatially flat spacetime), we find simple analytic expressions as follows:
\begin{equation}\label{eq:OXz}
\OX(z)=(1-\nu) f(z)-\Omo (1+z)^{3(1+\wm)}
\end{equation}
and
\begin{equation}\label{eq:OLz}
\OL(z)=\frac{\OLo-\nu}{1-\nu}+\nu f(z)\,,
\end{equation}
where we have defined
\begin{equation}\label{eq:deffz}
f(z)=\frac{\Omo(\wm-\wX)}{\wm-\wX+\epsilon}\,(1+z)^{3(1+\wm)}+g(z)
\end{equation}
and
\begin{equation}\label{eq:defgz}
g(z)=\left[\frac{1-\OLo}{1-\nu}-\frac{\Omo(\wm-\wX)}{\wm-\wX+\epsilon}\right]\ (1+z)^{3(1+\wX-\epsilon)}\,.
\end{equation}
In the above, $\wm=0,1/3$ for dust and radiation, respectively. Furthermore, we have defined the parameter combination
\begin{equation}\label{eq:epsilon}
\epsilon\equiv\,\nu\,(1+\wX)\,,
\end{equation}
which we will use as one of the basic fitting parameters of the $\CC$XCDM. As a second fitting parameter, we take $\wX$, 
the EoS of the cosmon. Finally,
for convenience, the following alternative combination of parameters is also introduced and will play the role of third fitting parameter:
\begin{equation}\label{eq:delta}
\delta\equiv\OXo(1+w_X)\,.
\end{equation}
From Eq.\,\eqref{eq:EOS}, we can see that its sign determines the effective (quintessence-like or phantom-like) behavior of the model around our time (cf. Sections\ \ref{sec:DataAnalysis} and \ref{sec:Discussion} for more details).
Altogether, $\CC$XCDM involves $n_{p, \CC{\rm XCDM}} - n_{p, \CC{\rm CDM}}=3$ additional free parameters $(\epsilon,w_X,\delta)$ as compared to the standard $\CC$CDM.
In Sec.\,\ref{sec:DataAnalysis}  we explain why we use these parameters, instead of e.g. $(\nu,\wX,\OXo)$, which is, of course, an equivalent triad. Whatever the choice, the $\CC$XCDM has the same number of extra d.o.f. as the aforementioned $w$XCDM model (although the parameters  are not identical in the two models).

The Hubble function of $\CC$XCDM can be straightforwardly computed as follows:
\begin{equation}\label{eq:FL2}
H^2(z)=H_0^2\,\left[\Omo\,(1+z)^{3(1+\wm)}+\OD(z)\,\right]\,,
\end{equation}
where
\begin{eqnarray}\label{eq:ODz}
\OD(z)=\frac{\OLo-\nu}{1-\nu}-\frac{\epsilon\,\Omo\,(1+z)^{3(1+\wm)}}{\wm-\wX+\epsilon}
+g(z)\,,\phantom{X}
\end{eqnarray}
with $g(z)$ as before. For $\nu=0$ and $\OXo=0$ one can easily check that the standard $\CC$CDM is exactly recovered.
Notice that for the $\CC$XCDM the following generalized cosmic sum rule holds:
\begin{equation}\label{eq:sumrule0}
\OMo+\ODo=\OMo+\OLo+\OXo=1\,.
\end{equation}
It follows that individually $\OLo$ and $\OXo$ can have any sign provided Eq.\,\eqref{eq:sumrule0} is preserved. It is  easy to see that for $w_X=-1$ (and any value of $\nu$) the $\CC$XCDM behaves as a DE model with constant $\OD=\OXo+\OLo$, although with $\OX$ and $\OL$ both evolving with the expansion:
\begin{equation}\label{eq:wxm1}
\begin{split}
&\OX(z)=\OXo+\nu\OMo\left[(1+z)^{3(1+\wm)}-1\right]\,,\\
&\OL(z)=\OLo+\nu\OMo\left[1-(1+z)^{3(1+\wm)}\right]\,.
\end{split}
\end{equation}
For $w_X=-1$ and $\nu=0$, $\CC$XCDM becomes an effective $\CC$CDM with $\OX=\OXo$ and $\OL=\OLo$ at all times and with a cosmological constant $\OD=\OXo+\OLo$. None of the last two situations is of particular interest to us, since they both behave $\CC$CDM-like. Another uninteresting situation would be $\wX\neq -1$ for $\nu=0$, as there is no running of the vacuum component, cf. Eq.\,\eqref{eq:OLz}.

From the former considerations, it is clear that only when the two characteristic features of the $\CC$XCDM are active simultaneously, namely when $\nu\neq 0$ and  $\wX\neq -1$ (and above all when $\wX<-1$) does the desired nontrivial behavior of the model reveal. In fact, for all the relevance of the cosmon component $X$, in the absence of running vacuum ($\nu=0$) the cosmon could not decay into VED and this would hinder the change of sign of $\rX$ and ultimately also prevent the composite DE fluid $\rD=\rL+\rX$ from performing the desired transition from ordinary phantom DE  (before crossing the phantom divide) into quintessence-like behavior (near our time). That is why the RVM must be the core structure of $\CC$XCDM.  We shall further discuss this crucial point quantitatively later on.

Let us now define what we shall call the coincidence ratio (related to the `cosmological coincidence problem'\, (cf. \cite{Peebles:2002gy} and references therein), as follows: $r\equiv \rD/\rho_m$. This ratio grows indefinitely in the $\CC$CDM since $\rD=\rv=$const. and $\rho_m\propto a^{-3}\to 0$ for $a\to \infty$. Thus, there is no reason why $r$ should have a value of order $1$ just at present (cosmic coincidence?). In the $\CC$XCDM, however, despite that we also have $\rho_m\propto a^{-3}\to 0$ for $a\gg1$,  $r$ remains bounded ($r<r_{\rm max}$) in a large part of the parameter space and for the entire expansion history, the bound being just of order one: $r_{\rm max}
={\cal O}(1)$. To show this remarkable property, which may be thought of as a possible solution or alleviation of the cosmic coincidence problem, we first compute $r$ explicitly in this model as a function of the cosmological redshift. We start by considering the radiation-dominated  epoch and in particular the BBN epoch. In this regime, $\wm=\wR\equiv1/3$ and $\rho_m(z)=\rho_r^0(1+z)^4$, and the coincidence ratio reads as follows:
\begin{equation}\label{eq:rBBN}
r_{\rm BBN}(z)
=\frac{\OLo-\nu}{(1-\nu)\,\ORo\,(1+z)^4}-\frac{\epsilon}{w_r-\wX+\epsilon}+R_N(z)\,,
\end{equation}
where we have defined the function
\begin{equation}\label{eq:RzM}
R_N(z)=\left[\frac{1-\OLo}{\ORo(1-\nu)}-\frac{\wR-\wX}{\wR-\wX+\epsilon}\right]\,(1+z)^{3\,(\wX-\epsilon)-1}\,.
\end{equation}
Clearly, this function  should not grow at high $z$ as otherwise the excessive amount of DE in the early stages would prevent standard BBN from occurring. This imposes the constraint $\wX-\epsilon<\frac13$, which ensures that $R_N(z)$ becomes residual at high energies, i.e. $R_N(z)\to 0$ for $z\to\infty$.  At these energies, the only term on the r.h.s. of Eq. \eqref{eq:rBBN} that survives is the constant contribution
\begin{equation}\label{eq:repsilon}
r_\epsilon\equiv -\frac{\epsilon}{w_r-\wX+\epsilon}\,.
\end{equation}
This term is restricted by BBN bounds to satisfy $|r_\epsilon|\lesssim 1\%$ \,(cf.\,\cite{Asimakis:2021yct,Cook:2025gra,Matei:2026zqv}, and references therein).

On the other hand, the coincidence ratio in the late universe (where $\wm=0$) reads
\begin{equation}\label{eq:rzMD}
r_{\rm}(z)
=\frac{\OLo-\nu}{(1-\nu)\,\Omo\,(1+z)^3}+\frac{\epsilon}{\wX-\epsilon}+R(z)\,,
\end{equation}
where
\begin{equation}\label{eq:RzM}
R(z)=\left[\frac{1-\OLo}{\Omo(1-\nu)}-\frac{\wX}{\wX-\epsilon}\right]\,(1+z)^{3\,(\wX-\epsilon)}\,.
\end{equation}

The interesting property of the $\CC$XCDM, first pointed out in \,\cite{Grande:2006nn}, is that the growth of $r(z)$ becomes bounded in the future. In fact, the redshift
of the maximum can be worked out from Eqs.\,(\ref{eq:rzMD})-\eqref{eq:RzM}:
\begin{equation}\label{zs}
1+\zm=\left[\frac{\OLo-\nu}{\wX\,(\OXo+\nu\,\OMo)-
\epsilon\,(1-\OLo)}\right]^{\frac{1}{3\,(1+\wX-\epsilon)}}\,.
\end{equation}
From the second
derivative of $r(z)$ at $z=\zm$,
\begin{equation}\label{eq:rpp}
r''(\zm)=\frac{9(1+\wX)}{(1+\zm)^5}\,\frac{\OLo-\nu}{\OMo}\,,
\end{equation}
we can see that a sufficient condition for the
extremum to be a maximum is that
 $(1+\wX)\,\left(\OLo-\nu\right)<0$.
This condition is fulfilled only for $\wX<-1$ and $\OLo>\nu$, or for $\wX>-1$ and $\OLo<\nu$. However, fitting the model to the data (cf. Sec. \ref{sec:DataAnalysis}) we find that $\wX<-1$, along with the condition $\OX^0<0$, which altogether corresponds to phantom matter (PM) behavior of the cosmon, see the Introduction. Thanks to that, the required type of crossing of the phantom divide obtains. Therefore, the PM option is definitely singled out by the type of crossing observed and offers a simultaneous alleviation of the cosmic coincidence problem (see Sec. \ref{sec:DataAnalysis} for numerical details).
The height  of the maximum ($r_{\rm
max}=r(\zm)$) determines the upper bound on the cosmic coincidence ratio in the $\CC$XCDM:
\begin{equation}\label{height}
r_{\rm
max}=\frac{1}{(1+\zm)^3}\,\frac{\OLo-\nu}{\OMo}\,\frac{1+\wX}{\wX-\epsilon}+\frac{\epsilon}{\wX-\epsilon}\,.
\end{equation}
The numerical analysis in Sec. \ref{sec:DataAnalysis} demonstrates that, in the region of parameter space where the crossing occurs,  $r_{\rm max}
={\cal O}(1)$.
One can also show that the maximum must lie ahead our time, as otherwise the acceleration condition $\ddot a>0$ of the universe in the neighborhood of our past would be violated.

After surpassing the point $z=\zm$ into the future, the ratio (\ref{eq:rzMD}) drops relatively fast (in redshift units) to $r(z_s)=-1$ since $\OD(z_s)=-\Omo\,(1+z_s)^3$, where $z_s$ is the `stopping' -- or, more accurately, the return -- point, where $H(z_s)=0$, see Eq.\,\eqref{eq:FL2}. From this point onward, the universe heads back towards the Big Crunch (see the plot of the cosmic coincidence ratio $r$ in the next section obtained from our numerical fit to the data).

Let us finally come back to the effective EoS of the $\CC$XCDM, $\weff$, which is the central focus of our work. It becomes fully determined as a function of the cosmic redshift by substituting the above given expressions for $\OX(z)$ and $\OD(z)$  into Eq.\,\eqref{eq:EOS}. The crossing of the phantom divide occurs for the value $z=z^*$ where $\OX(z^*)=0$, as then $\weff(z^*)=-1$.
Since the sign of $(1+\wX)\OX(z)$ is crucial to determine the EoS behavior of the total dark energy $\rD$ in the $\CC$XCDM, let us further work out Eq.\,\eqref{eq:OXz} as follows:
\begin{equation}\label{eq:OXbis}
\begin{split}
\OX(z)=&\left(\OXo+\frac{\nu (1+\wm)\ \Omo}{\wm-\wX+\epsilon}\right) (1+z)^{3(1+\wX-\epsilon)}\\
&-\frac{\nu\,(1+\omega_m)\ \Omo}{\wm-\wX+\epsilon} (1+z)^{3(1+\wm)}\,,
\end{split}
\end{equation}
where we have made explicit use of the generalized cosmic sum rule \eqref{eq:sumrule0}. Notice the correct normalization $\OX(0)=\OXo$. The more explicit formula \eqref{eq:OXbis} is helpful in better understanding the behavior of $\OX(z)$ at low and high redshift. As we have just seen, at low redshift $\OX(z)\simeq\OXo$ ($z\gtrsim0$) and hence under PM conditions for the component $X$ (which entail $\OX(z)<0$, also necessary for alleviating cosmic coincidence), the effective EoS \eqref{eq:EOS} of the composite fluid behaves as quintessence near our time ($\weff\gtrsim-1$). On the other hand, from the fact that $1+\wX-\epsilon<0$ (due to $\wX<-1$ and $\epsilon>0$) it follows that the first term on the r.h.s. of Eq.\,\eqref{eq:OXbis} goes to zero for $z\gg 1$. In particular, for relatively large $z$ but still in the late universe ($\wm=0$), we have
\begin{equation}\label{eq:OXbis2}
\OX(z\gg 1)\simeq
\frac{\nu\ \Omo}{\wX-\epsilon} (1+z)^{3}>0\,,
\end{equation}
because $\nu<0$ (ensuring the decay of $X$ into vacuum) and the fact that the denominator is negative. In fact, from the fitting results we will confirm numerically that this holds even for $z\gtrsim 1$. One can also show from \eqref{eq:OLz} that
\begin{equation}\label{eq:OLambda}
\OL(z\gg 1)\simeq\
\frac{\nu\,\wX}{\wX-\epsilon} \ \Omo (1+z)^{3}<0\,,
\end{equation}
while around the current universe ($z\gtrsim 0$)  we have $\OL(z)>0$ in the region targeted by our fit.
Overall, for $X$  in the PM region, we find that at relatively large $z$, we have $\OX(z)>0$, $\OL(z)<0$ and $(1+\wX)\OX(z)<0$,  and hence  $\weff$ behaves as phantom DE, while in the current universe we have $\OX(z)<0$, $\OL(z)>0$ and  $(1+\wX)\OX(z)>0$, and as a result $\weff$ behaves as quintessence. Ergo, there must exist a crossing point $z^{*}$ of the phantom divide between the two regimes.  The value of $z^*$ around our time ($\wm=0$)  can be explicitly computed from Eq. \eqref{eq:OXbis}:
 \begin{equation}\label{eq:zstar}
1+z^{*}=\left\{\frac{\nu\,\OMo}{\nu\,\OMo+\OXo (\epsilon-\wX)}\right\}^{-\frac{1}{3 (\epsilon-\wX)}}\,.
\end{equation}
From this result, we confirm once more in a very manifest way that the two characteristic features of the $\CC$XCDM must be alive simultaneously. For, there would be no crossing of the phantom divide in the  absence of running vacuum (i.e. for $\nu=0$); and for $\OXo=0$ the crossing lies at $z=0$, which is not what we observe. Finally,  for $\wX=-1$ there cannot be crossing either, since $w_{\rm eff}=-1$ $\forall z$, see Eq.\,\eqref{eq:EOS}. In general, one can prove that in the relevant region of the $\CC$XCDM parameter space all possible crossings are one way only, viz. from phantom DE ($\weff<-1$) into quintessence ($\weff\gtrsim -1$), this being the way observed by DESI. From the numerical results presented in the next section (cf. Table \ref{tab:results}), we find that the most favored crossings (at 95\% CL) occur within the approximate redshift interval $z^{*}\simeq 0.2-0.9$, which, interestingly enough, fits in with the approximate range highlighted by DESI and occur in a domain of parameter space where there is  a substantial alleviation of the cosmic coincidence problem.

It goes without saying that phantom matter plays a momentous role in this whole story. To categorize the nature of PM and better understand its meaning in the context of the EoS diagram of cosmic fluids, see Fig.\,1 of \cite{Gomez-Valent:2024tdb}. In it, we can see that PM is in the antipodes of the usual phantom DE, since  in contrast to the latter, PM injects positive pressure ($P_X>-\rho_X>0$) at the expense of negative energy density ($\rX<0$). Thanks to this feature encoded in the cosmon decay into vacuum, the observed crossing of the phantom divide performed by the effective EoS $\weff$ can be efficiently  realized in a large region of the $\CC$XCDM parameter space and without further specifying the nature of $X$.

In Sec.\,\ref{sec:Discussion},  we shall retake in more detail the analysis of the phenomenological consequences of the $\CC$XCDM model and confirm the encouraging conclusions derived from the above preliminary discussion. Needless to say, the rigorous results can only be secured  after the model is fitted against the observational data. This is performed in the next section.


\section{Data and numerical analysis}\label{sec:DataAnalysis}

In this work, we constrain the $\Lambda$XCDM as well as the $\Lambda$CDM and CPL models using the following three data sets:

\begin{itemize}
\item {\it CMB:} The Planck temperature
(TT), polarization (EE) and cross (TE) power spectra. More concretely, we make
use of the \texttt{simall}, \texttt{Commander} and \texttt{NPIPE PR4} \texttt{CamSpec} likelihoods \citep{Efstathiou:2019mdh, Rosenberg:2022sdy} for
$\ell<30$ and $\ell\geq 30$, respectively, and the \texttt{NPIPE PR4} CMB
lensing likelihood \citep{Carron:2022eyg}.

\item {\it BAO:} The BAO data from DESI Data Release 2
(DR2). See Table
IV in \cite{DESI:2025zgx}.

\item {\it SNIa:} We combine the CMB+BAO data with two different SNIa compilations, considered separately: Pantheon+ \citep{Brout:2022vxf} and DES-Dovekie \citep{DES:2025sig}. The latter supersedes the older DES-Y5 sample \citep{DES:2024hip,DES:2024jxu}, which was affected by calibration systematics. DES-Dovekie has been found to strengthen the evidence for DDE within the CPL parametrization compared to Pantheon+. Here, we examine whether this result is also supported in the context of the $\Lambda$XCDM model. To distinguish these two combinations of data, we refer to them as CMB+BAO+SNIa(Pan) and CMB+BAO+SNIa(Dov), respectively. With Pantheon+ we sample over the absolute magnitude $M$, whereas with DES-Dovekie it is marginalized over, following the approach adopted in the official likelihood of the DES Collaboration.

\end{itemize}

It is worth mentioning that additional BBN priors could be included. In particular, a viable $\Lambda$XCDM cosmology should not magnify the DE to matter ratio during nucleosynthesis, i.e. $r_{\rm BBN}$ from Eq. \eqref{eq:rBBN}. We have checked that for parameter values preferred by the data, the coincidence ratio lies in the expected ballpark of $\sim1\%$ (\jtext{as mentioned}  in the previous section), so we do not impose any extra BBN prior.

In this analysis, we do not use large-scale structure formation data nor the SH0ES calibration of $H_0$ since we want to better compare with DESI results and also with previous analyzes existing in the literature, where only the above datasets are used in the main. Our focus here is to demonstrate that the $\CC$XCDM model can help in a very efficient way to cross the phantom divide in the precise manner observed by DESI.  The effect of $\CC$XCDM on the $H_0$ tension will not be addressed here, as it would require a more detailed study
(see Sec.\,\ref{sec:Discussion}).

We have implemented the $\Lambda$XCDM model in the Einstein--Boltzmann code \texttt{CLASS} \citep{Lesgourgues:2011re,Blas:2011rf}, which we use to solve the cosmological background and linear perturbation equations and to compute the expansion history, cosmological distances, and CMB power spectra entering our likelihood. The  perturbation equations are implemented along the same lines as in \cite{deCruzPerez:2025dni}. We assume that DE does not cluster, setting its sound speed to 1 (cf. \cite{Grande:2008re} for alternative scenarios), and use the minimal normal hierarchy configuration for neutrinos, with only one massive neutrino of mass 0.06 eV. We run Monte Carlo Markov chains (MCMC) using the Metropolis-Hastings algorithm \citep{Metropolis:1953am,Hastings:1970aa} with the aid of \texttt{Cobaya} \citep{Torrado:2020dgo} and the resulting chains are analyzed with \texttt{GetDist} \citep{Lewis:2019xzd}.  We consider the Gelman-Rubin convergence criterion \citep{GelmanRubin1992} $R-1=0.02$ for $\Lambda$CDM and CPL, whereas for $\Lambda$XCDM we relax it to $R-1=0.03$, as its convergence is significantly slower owing to the presence of non-Gaussian features in the posterior distribution.

In our MCMC analysis of $\Lambda$XCDM we allow the three additional parameters $(\epsilon,w_X,\delta)$ (cf. Eqs.~\eqref{eq:epsilon}-\eqref{eq:delta}) to vary freely, together with the six standard cosmological parameters of the $\Lambda$CDM model, namely $\omega_{\rm cdm}$, $\omega_b$, $H_0$, $A_s$, $n_s$, and $\tau_{\rm reio}$. For the CPL, instead, we vary $w_0$ and $w_a$ alongside the $\Lambda$CDM parameters. An a priori natural choice for the additional free parameters of the $\CC$XCDM would perhaps be to vary $(\nu,w_X,\Omega_X^0)$, as they are more fundamental. However, in practice it is not so convenient. As already discussed in Sec.~\ref{sec:CompositeDE}, when $w_X=-1$ the $\Lambda$XCDM model is indistinguishable from the standard $\Lambda$CDM model, even if $\nu\neq0$, i.e., even in the presence of an energy transfer between the vacuum and the cosmon. This inner transfer affects individually the two components of the DE,  but  \jtext{in that case} the total DE remains constant,  see Eq.\,\eqref{eq:wxm1}, and its contribution to the Friedmann and acceleration equations is identical to that of a \jtext{cosmological constant $\rD=$const.}, independently of the value of $\Omega_X^0$. Consequently, in the parameter space spanned by the triad $(\nu,w_X,\Omega_X^0)$, the standard $\Lambda$CDM model is represented not by a single point but by an entire line. This introduces a strong degeneracy among these parameters -- corresponding to lines of $\delta={\rm const}.$ and $\epsilon=$const. --, as we have explicitly verified by also performing MCMC runs using this parametrization. By contrast, when using $(\epsilon,w_X,\delta)$, the $\Lambda$CDM limit is recovered only at the single point $(\epsilon,w_X,\delta)=(0,-1,0)$ of parameter space, thereby removing that degeneracy and leading to a more efficient numerical exploration of the parameter space.  As we will see in Sec. \ref{sec:Discussion}, the likelihood already disfavors values of $w_X$ close to $-1$, but to avoid unreasonably large values of $\OXo$ and $\nu$ when the sampler explores those regions of parameter space we impose external priors, $|\delta/(1+w_X)|<0.98$ and $|\epsilon/(1+w_X)|<0.5$ (or, equivalently, $|\Omega_X^0|<0.98$ and $|\nu|<0.5$).

The three parameters $(\epsilon,w_X,\delta)$ are therefore of practical use, and encode relevant physical information of the model. As noted in Sec. \ref{sec:CompositeDE}, the sign of $\delta$ determines the current quintessence- or phantom-like nature of the composite DE fluid. In the next section, we will show that current data prefer $\delta>0$ and, hence, quintessence in the late universe, in full consistency with the values of $w_0>-1$ obtained within the CPL parametrization.

Regarding $\epsilon$, it differs from zero only if the vacuum exchanges energy with the cosmon (i.e., if $\nu\ne 0$) and this interaction contributes genuinely to the dynamics of the composite DE, which can occur only if $w_X\ne -1$. Recall that if $w_X\ne -1$ but $\nu=0$, the composite DE is still dynamical, since $\dot{\rho}_D=\dot{\rho}_X\ne 0$; however, in this case the vacuum does not contribute to the dynamics. If, instead, $w_X=-1$ and $\nu\ne 0$, then $\rho_D=\mathrm{const}$, \jtext{as previously noted}. The crossing of the phantom divide can occur only if the composite DE is dynamical and if $\nu \neq 0$ and negative, as clearly shown by Eqs. \eqref{eq:EOS} and \eqref{eq:zstar} -- \jtext{notice that $\epsilon-\wX>0$ in the region of interest (singled out by our fit, cf. Table \ref{tab:results}) and therefore  the crossing is in the past ($z^*>0)$}. Therefore, if $w_X<-1$ (as required to alleviate the cosmic coincidence problem), we expect the evidence for such a crossing found within the CPL parametrization to translate into a preference for $\epsilon>0$, and hence $\nu<0$. Thus, although having $w_X \neq -1$ in the phantom region is necessary to allow for an effective DE dynamics, it is not sufficient by itself to explain the current cosmological data.

We note that the fitting analysis using the simplest parameterization $w$CDM of the dynamical DE\,\citep{Turner:1997npq} -- obtained from the more complete CPL parametrization by setting $w_a=0$ and, hence, assuming a constant (but non-zero) EoS parameter -- does not reveal any signature of new physics. Instead, it yields values of $w_0$ that are fully compatible with $-1$ at the $68\%$ CL (see~\cite{DESI:2024mwx,Gomez-Valent:2024ejh,DESI:2025zgx, Gonzalez-Fuentes:2025lei,Gonzalez-Fuentes:2026rgu}), essentially because it does not exhibit phantom-divide crossing. Therefore, any preference for physics beyond $\Lambda$CDM within the $\Lambda$XCDM model must be accompanied by a preference for non-zero (and positive) values of $\epsilon$ and $\delta$, and  $1+w_X<0$.

We compare the fitting performance of the $\Lambda$XCDM with that offered by the $\Lambda$CDM and CPL using two alternative statistical methods that duly penalize the use of additional parameters -- note that CPL and $\Lambda$XCDM have two and three more parameters than the standard model, respectively. On the one hand, we employ the traditional Akaike (AIC) \citep{Akaike} information criterion, which is defined as

\begin{equation}\label{eq:AIC}
    {\rm AIC} =  \chi^2_{\rm min}+2n_p
\end{equation}
when the number of data points entering the fit is much larger than the number of free parameters of the model, $n_p$, as it is in the case under study. A positive difference $\Delta {\rm AIC}=$AIC$_{\Lambda{\rm CDM}}-$AIC$_{i}$ implies that the model $i$ performs better than the $\Lambda$CDM, with values $2 \leq \Delta\textrm{AIC} < 6$ usually considered to show \textit{positive evidence} according to Jeffreys' scale, values in the range $6 \leq \Delta\textrm{AIC} < 10$ indicating \textit{strong evidence}, and values $\Delta\textrm{AIC}>10$ pointing to a  \textit{very strong} statistical evidence supporting the model $i$ against the standard $\CC$CDM.

\begin{table*}[t!]
   \centering
\resizebox{\textwidth}{!}{
   \begin{tabular}{c|ccc|ccc}
   \hline
    & \multicolumn{3}{c|}{\textbf{CMB + DESI DR2 + PantheonPlus}} & \multicolumn{3}{|c}{\textbf{CMB + DESI DR2 + DES Dovekie}} \\
   \hline \hline
   \textbf{Parameter} & $\mathbf{\Lambda}$\textbf{CDM} & \textbf{CPL} & $\mathbf{\Lambda}$\textbf{XCDM} & $\mathbf{\Lambda}$\textbf{CDM} & \textbf{CPL} & $\mathbf{\Lambda}$\textbf{XCDM} \\
   \hline
   $H_0 \, \left[ \mathrm{km/s/Mpc} \right]$ & $68.06\pm 0.29$ & $67.53\pm 0.59$ & $67.07\pm 0.57$ & $68.04\pm 0.28$ & $67.33\pm 0.55$ & $66.68\pm 0.59$\\
   $10^2 \omega_{\rm b}$ & $2.231\pm 0.012$ & $2.222\pm 0.013$ & $2.227\pm 0.012$ & $2.230\pm 0.012$ & $2.221\pm 0.013$ & $2.227\pm 0.012$\\
   $10 \omega_{\rm cdm}$ & $1.1774\pm 0.0062$ & $1.1887\pm 0.0083$ & $1.1669\pm 0.0067$ & $1.1780\pm 0.0061$ & $1.1904\pm 0.0084$ & $1.1663\pm 0.0066$\\
   $\ln(10^{10}A_s)$ & $3.046\pm 0.014$ & $3.037\pm 0.014$ & $3.028\pm 0.015$ &$3.046^{+0.013}_{-0.014}$ & $3.036\pm 0.014$ & $3.027^{+0.016}_{-0.014}$ \\
   $n_s$ & $0.9677\pm 0.0034$ & $0.9649\pm 0.0038$ & $0.9667\pm 0.0035$ & $0.9677\pm 0.0034$ & $0.9646\pm 0.0036$ & $0.9669\pm 0.0035$\\
   $\tau_{\rm reio}$ & $0.0586^{+0.0065}_{-0.0073}$ & $0.0535\pm 0.0070$ & $0.0492\pm 0.0074$ & $0.0585^{+0.0063}_{-0.0072}$ & $0.0527\pm 0.0072$& $0.0489^{+0.0081}_{-0.0070}$\\
   $M \, \left[ \mathrm{mag.} \right]$ & $-19.4201\pm 0.0089$ & $-19.419\pm 0.014$ & $-19.435\pm 0.012$ & $-$ & $-$ & $-$\\
   $w_0$ & $-$ & $-0.844\pm 0.054$ & $-$ & $-$ & $-0.809\pm 0.056$ & $-$\\
   $w_a$ & $-$ & $-0.59^{+0.22}_{-0.19}$ & $-$ & $-$ & $-0.71^{+0.23}_{-0.21}$ & $-$ \\
   $w_X$ & $-$ & $-$ & $< -1.66\; (95\%)$ & $-$ & $-$ & $< -1.96\; (95\%)$\\
   $\epsilon\equiv \nu(1+w_X)$ & $-$ & $-$ & $0.0239^{+0.008}_{-0.010}$ & $-$ & $-$ & $0.0259^{+0.009}_{-0.010}$ \\
   $\delta\equiv \Omega_X^0(1+w_X)$ & $-$ & $-$ & $0.107^{+0.047}_{-0.063}$ & $-$ & $0.3131\pm 0.0054$ & $0.152^{+0.061}_{-0.073}$ \\
   \hline
   $\Omega_{\rm m}^0$ & $0.3033\pm 0.0037$ & $0.3109\pm 0.0057$ & $0.3100\pm 0.0056$ & $0.3037\pm 0.0036$ & $0.3131\pm 0.0054$ & $0.3134\pm 0.0058$\\
   $r_{\rm d} \, \left[ \mathrm{Mpc} \right]$ & $147.78\pm 0.19$ & $147.57\pm 0.21$ & $148.67\pm 0.32$ & $147.77\pm 0.19$ & $147.53\pm 0.21$ & $148.70\pm 0.32$\\
   $\sigma_{12}$ & $0.7935\pm 0.0062$ & $0.8006\pm 0.0068$ & $0.8091\pm 0.0081$ & $0.7938\pm 0.0060$ & $0.8015\pm 0.0068$ & $0.8090\pm 0.0082$\\
   $S_8$ & $0.8093\pm 0.0078$ & $0.8221\pm 0.0088$ & $0.8258\pm 0.0091$ & $0.8099\pm 0.0076$ & $0.8244\pm 0.0088$ & $0.8270\pm 0.0093$\\
   $\nu$ & $-$ & $-$ & $-0.0134^{+0.0061}_{-0.0026}$ & $-$ & $-$ & $-0.0128^{+0.0052}_{-0.0030}$\\
   $\Omega_X^0$ & $-$ & $-$ & $-0.059^{+0.033}_{-0.020}$ & $-$ & $-$ & $-0.072^{+0.029}_{-0.024}$\\
   $z^*$ & $-$ & $0.38^{+0.22}_{-0.20}\; (95\%)$ & $0.54^{+0.44}_{-0.31}\; (95\%)$ & $-$ & $0.38^{+0.17}_{-0.15}\; (95\%)$ & $0.55^{+0.33}_{-0.24}\; (95\%)$ \\
   \hline
   $\chi^2_{\rm min}$ & $12393.89$ & $12385.64$ & $12381.87$ & $12622.30$ & $12611.22$ & $12608.36$ \\

   $\Delta\mathrm{AIC}$  & $-$ & $4.25$ & $6.02$ & $-$ & $7.08$ & $7.93$ \\
   $p$-value & $-$ & $0.0162$ & $0.00730$ & $-$ & $0.00393$ & $0.00300$ \\
   $E_{\Lambda\mathrm{CDM}}$ & $-$ & $2.41\sigma$ & $2.68\sigma$ & $-$ & $2.88\sigma$ & $2.97\sigma$\\
   \hline
   \end{tabular}}
   \caption{Means with $68\%$ CL values and statistics related to $\chi^2_{\rm min}$ for the 2 data combinations and models $\Lambda$CDM, CPL and $\Lambda$XCDM. The difference of Akaike information criteria is computed with respect to $\Lambda$CDM, $\Delta\mathrm{AIC} \equiv \mathrm{AIC}_{\Lambda\mathrm{CDM}} - \mathrm{AIC}_\mathrm{model}$, with $n_{p,\CC\mathrm{XCDM}} - n_{p,\CC\mathrm{CDM}}=3$. For $\Lambda$XCDM, $w_X$ and $z^*$ uncertainties are at $95\%$ due to their non-Gaussian features.  }
   \label{tab:results}
\end{table*}

On the other hand, we employ the likelihood-ratio test \citep{NeymanPearson1933} to estimate the exclusion level of $\Lambda$CDM with respect to the $\Lambda$XCDM and CPL models, assuming the validity of Wilks' theorem in the models under study \citep{Wilks1938}, which must be considered only as a first approximation due to the presence of non-gaussianities in the parameter posteriors. We basically follow the methodology of, e.g., \cite{DESI:2025zgx}. Under the null hypothesis that $\Lambda$CDM is the true cosmological model, the difference between the minimum chi-square values in the standard model and a nested model $i$,
\begin{equation}
\Delta\chi^2_{\rm min}
=
\chi^2_{{\rm min},\Lambda{\rm CDM}}
-
\chi^2_{{\rm min},i},
\end{equation}
follows a $\chi^2_\zeta$ distribution with
\begin{equation}
\zeta = n_{p,i}-n_{p,\Lambda{\rm CDM}}
\end{equation}
degrees of freedom. We compute the $p$-value associated with the observed value of $\Delta\chi^2_{\rm min}$ and the confidence level at which the null hypothesis can be rejected. As in \cite{DESI:2025zgx}, we express the resulting $p$-values in terms of the equivalent Gaussian significance, $\xi$, by solving
\begin{equation}
p\text{-value}
=
1-\frac{1}{\sqrt{2\pi}}
\int_{-\xi}^{\xi}
e^{-y^2/2}\,dy\,.
\label{eq:pvalue_sigma}
\end{equation}
The exclusion level of $\Lambda$CDM is simply given at $E_{\Lambda{\rm CDM}}=\xi\sigma$ CL. In the next section, we will show that both the AIC and $E_{\Lambda{\rm CDM}}$ provide consistent assessments of the statistical preference for the model, so their combined use reinforces our conclusions.

Finally, we note that, in the next section, we also study the profile likelihood of $w_X$ to assess the sensitivity of the data to this parameter and the impact of potential degeneracies in the model. We construct the corresponding profile likelihoods directly from the Markov chains following the method used in \cite{Gomez-Valent:2022hkb}. We essentially bin $w_X$ and search for the minimum value of $\chi^2$ in each bin. The corresponding values of the remaining parameters are then recorded and used to produce the curves shown in Figs. \ref{fig:profile_zcross}--\ref{fig:ratio}. As it will become clear from this analysis, one is forced to choose prior bounds for $w_X$, which we take to be flat and in the range $-4\leq w_X\leq -1$. For the remaining parameters, we also use flat priors, much broader than the posteriors.

\section{Discussion}\label{sec:Discussion}

Our fitting results obtained with the two data combinations described in the preceding section are presented in Table \ref{tab:results}. In Appendix \ref{sec:A}, we provide additional information, including the breakdown of the $\chi^2_{\rm min}$ contributions in Table \ref{tab:table_chi2} and the corresponding contour plots shown in Fig. \ref{fig:full_triangle_plot}. These supplementary results will facilitate a more detailed discussion of our findings. While the results obtained for the $\Lambda$CDM and CPL models are already well established (see, e.g., \cite{DESI:2025zgx,DES:2025sig} for analyzes including CMB data from ACT and SPT, and \cite{Gonzalez-Fuentes:2026rgu} for an analysis based on a data set closely related to ours), we deem it important to use them as baseline models. In particular, it has become common practice in the literature to use the CPL parametrization as a phenomenological  benchmark for DDE models capable of crossing the phantom divide.

The first important outcome that emerges from Table~\ref{tab:results} is the remarkable fitting performance of the $\Lambda$XCDM model. It provides a significantly improved fit to the current data, yielding minimum chi-squared values, $\chi^2_{\rm min}$, that are approximately $3\!-\!4$ units lower than those obtained with the CPL parametrization and $12\!-\!14$ units lower than those of $\Lambda$CDM. The improvement affects almost all individual data sets, but it is most prominently observed in the description of the high-$\ell$ CMB data from Planck (see Table \ref{tab:table_chi2} and also Figs. \ref{fig:dist_dove}-\ref{fig:dist_PP} in Appendix \ref{sec:B}). Thus, $\Lambda$XCDM is able to outperform the CPL model. At first sight, this result may not appear surprising, since $\Lambda$XCDM has one additional free parameter compared to CPL. However, we find that it remains favored even after accounting for the penalty associated with the extra parameter. Indeed, within the $\Lambda$XCDM framework, $\Lambda$CDM is excluded at slightly higher significance than in the CPL case, namely at $2.68\sigma$ and $2.97\sigma$ using the CMB+BAO+SNIa(Pan) and CMB+BAO+SNIa(Dov) data sets, respectively, compared with $2.41\sigma$ and $2.88\sigma$ for the CPL parametrization. The evidence for DDE increases with DES-Dovekie, also in the context of $\Lambda$XCDM. Importantly, the same conclusions hold when we perform the model comparison using the AIC instead of the likelihood-ratio test. We find strong evidence in favor of the $\Lambda$XCDM model for the two SNIa samples considered, with $\Delta$AIC values ranging from $\sim 6$ for PantheonPlus to $\sim 8$ for DES-Dovekie. These results again indicate a better performance of $\Lambda$XCDM compared to the CPL model.

\begin{figure}[t!]
    \centering
    \includegraphics[scale=0.4]{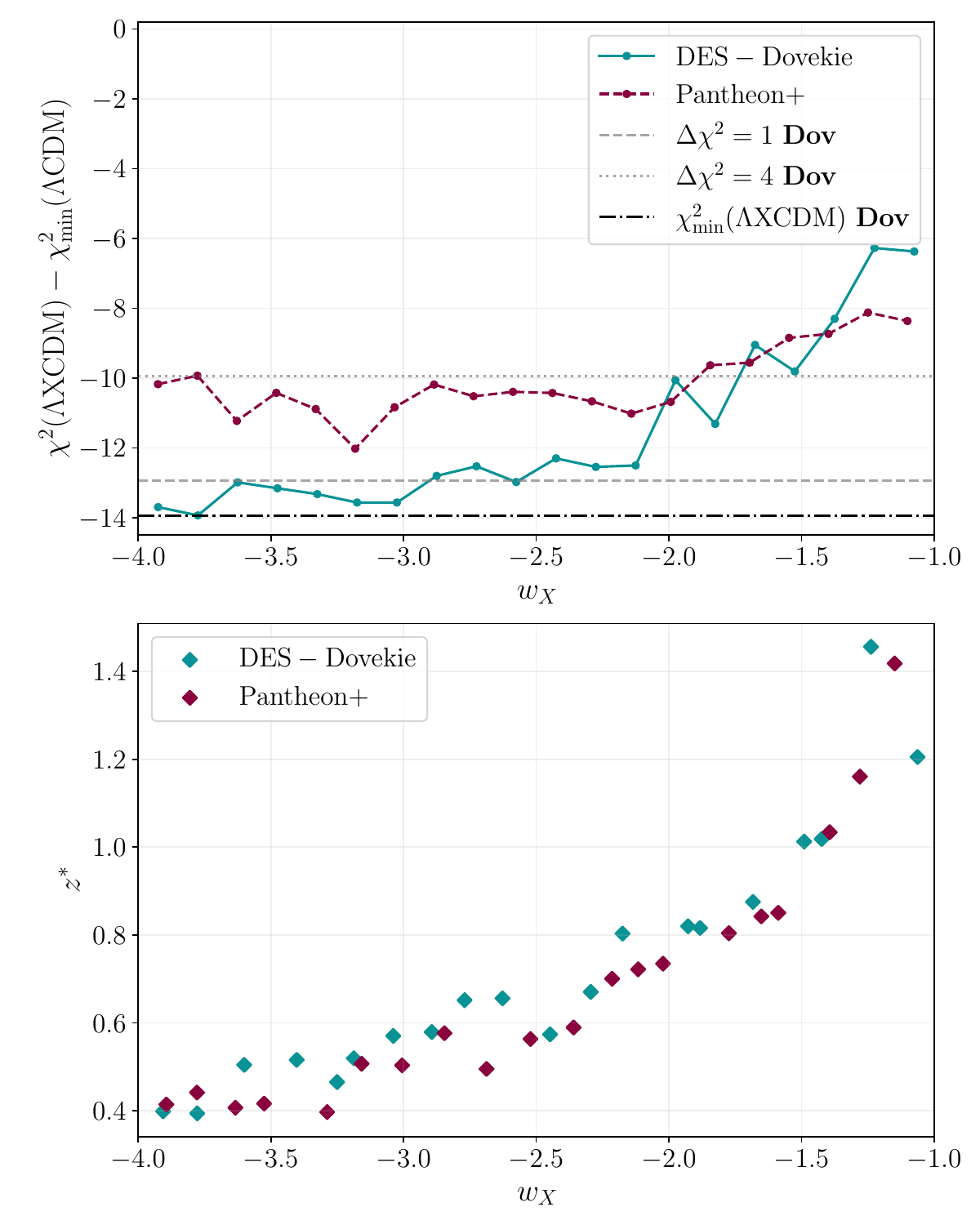} \caption{{\it Upper:} Profile likelihood in $w_X$ with respect to the minimum $\chi^2$ point of $\Lambda$CDM for each of the 2 datasets. {\it Lower:} Crossing redshift of the phantom divide for the profiled values. Note that in the upper plot we display the central bin values and in the lower one the \jtext{actual values of}  $w_X$ that produce the minimum within that bin.}
    \label{fig:profile_zcross}
\end{figure}

As anticipated in the previous sections, we find a preference for values of $w_X$ in the phantom region, i.e., $w_X<-1$, and for positive values of $\delta$ and $\epsilon$, which in turn imply negative values of $\Omega_X^0$ and $\nu$. Our results indicate that data prefer that energy flows from the cosmon -- which behaves as phantom matter in the late universe -- into the vacuum. Curiously, current data are not capable of placing tight constraints on $w_X$, but rather impose an upper bound: $w_X<-1.66$ with Pantheon+ and $w_X<-1.96$ with DES-Dovekie, both at $95\%$ CL. The latter shifts the posterior distribution toward even more negative values. Despite the fact that the one-dimensional marginalized posterior distributions of $w_X$ seem to favor the region of  smallest values of $w_X$ allowed by the prior (cf. Fig. \ref{fig:full_triangle_plot}) -- notice that the posteriors are monotonic decreasing and are cut by the lower bound of the prior --, the analysis of the profile likelihood for this parameter reveals that this conclusion is just an artifact, caused by volume effects in the MCMC analysis. In fact, the upper plot of Fig. \ref{fig:profile_zcross} shows that the minimum value of $\chi^2$ in each bin of $w_X$ remains stable for values of $w_X<-2$, producing a plateau. While the upper bounds reported in Table \ref{tab:results} may change under a different choice of the prior on $w_X$, the shape of the profile likelihood remains stable under such variations. This indicates that the data cannot efficiently discriminate between different values of $w_X$ in the phantom regime, provided that $w_X$ remains sufficiently below $-1$.  This further justifies the necessity of imposing the otherwise arbitrary prior bounds on $w_X$ to efficiently explore the viable parameter space for $\Lambda$XCDM.

\begin{figure}[t!]
    \centering
    \includegraphics[scale=0.5]{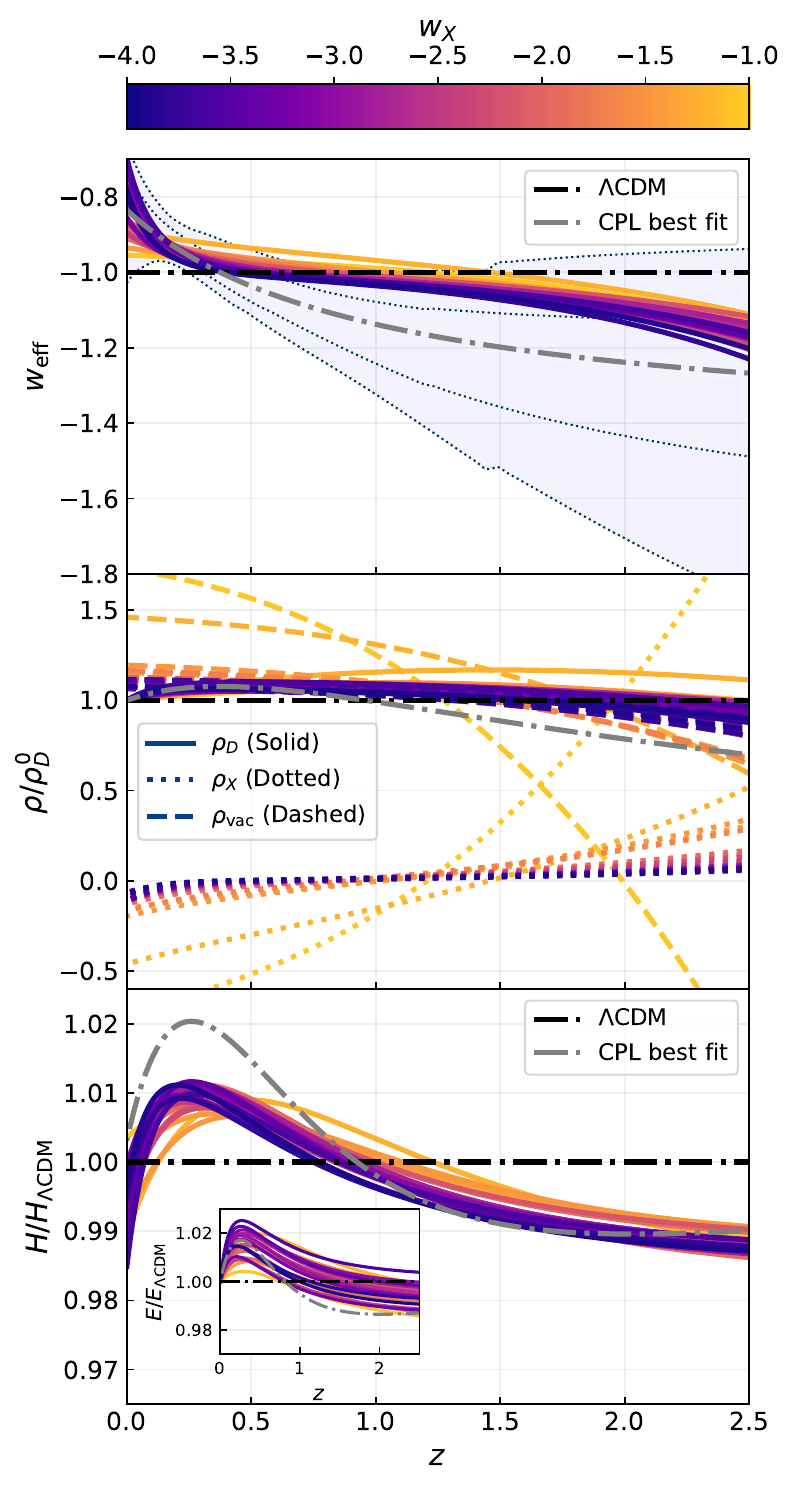} \caption{Curves for each of the profiled values of $w_X$ with CMB+BAO+SNIa(Dov). From top to bottom we display: i) the effective EoS \jtext{of the $\CC$XCDM and CPL} ($w_{\rm eff}$) with the 68\% and 95\% CL bands from \cite{Gonzalez-Fuentes:2026rgu}; ii) the total and individual DE components of the $\CC$XCDM, viz. $\rho_D$ and $(\rho_X, \rho_{\rm vac})$ normalized with respect to $\rho_D^0$, \jtext{including the DE density of the CPL}; iii) the Hubble rate $H(z)$ \jtext{of these models} with respect to the fiducial {\it Planck} PR4 $\Lambda$CDM with $H_0 = 67.26\,{\rm km/s/Mpc}$ and $\Omega_{\rm m}^0=0.315$ \citep{Rosenberg:2022sdy}. In the inner subplot the analogous figure for $E\equiv H/H_0$ is displayed. }
    \label{fig:tripleplot}
\end{figure}

Notably, all the points used to construct the profile likelihood lead to a crossing of the phantom divide, but the crossing does not occur at the same cosmic epoch. In the lower panel of Fig. \ref{fig:profile_zcross}, we show the redshift of the crossing, $z^*$ (Eq. \ref{eq:zstar}), as a function of the profiled value of $w_X$. We find a positive correlation between $w_X$ and $z^*$. As expected, the values of $w_X$ that provide a better description of the cosmological data, namely those with $w_X<-2$, yield a crossing redshift in the range $z^*\sim 0.4-0.8$, consistent with the values inferred from model-agnostic reconstruction analyzes \citep{Gonzalez-Fuentes:2025lei}. This can be also clearly seen in the upper panel of Fig. \ref{fig:tripleplot} and in the derived values of $z^*$ in Table \ref{tab:results}. Current data is therefore unable to provide tight constraints on the cosmon EoS parameter and a good fit can be achieved by a wide range of values of $w_X<-1$, which might be a reflection of the fact that dark energy microphysics may be affected by permanent underdetermination \citep{Wolf:2023uno}. Notwithstanding this, the existence of a wide parameter space for $\Lambda$XCDM that can produce phantom crossing is highly remarkable and implies that a plethora of models effectively described by a constant $w_X$ can give rise to such dynamics. In other words, there seems to be  a kind of `universality' in the crossing feature of the phantom divide by the $\CC$XCDM over a large family of cosmological models, each contributing with particular realizations of the cosmon $X$.

\begin{figure}[t!]
    \centering
    \includegraphics[scale=0.4]{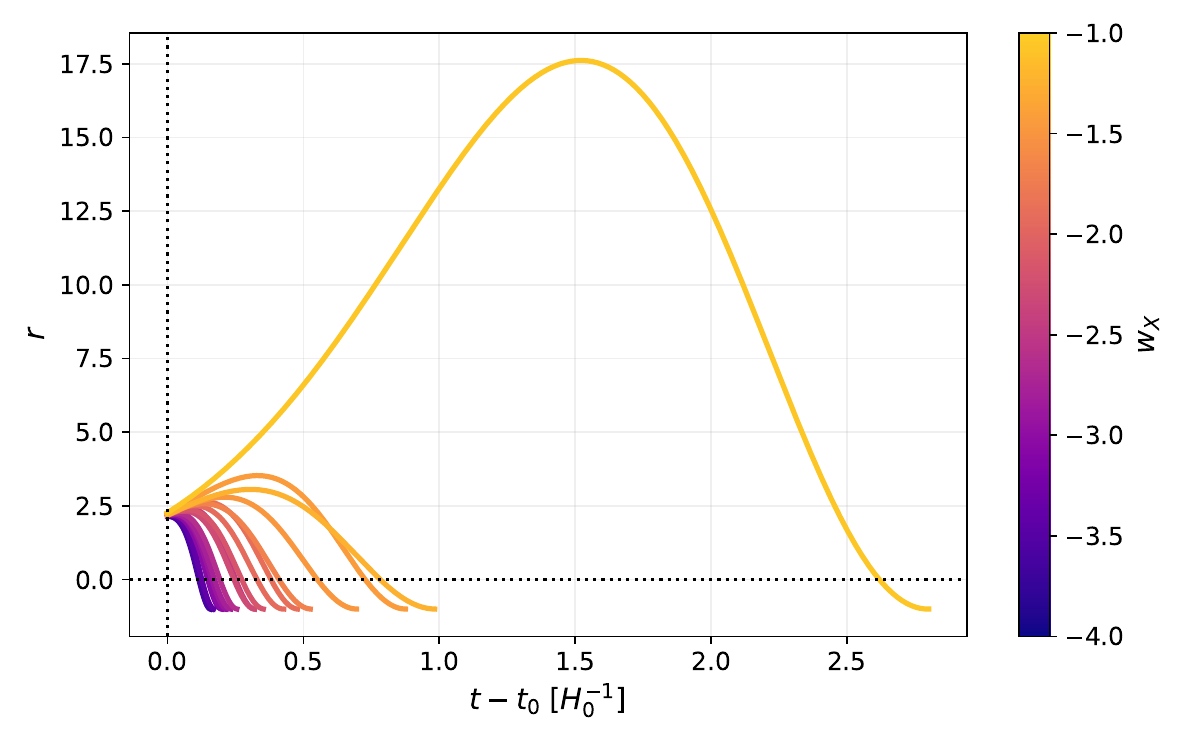} \caption{Coincidence ratio for the profile values of $w_X$ with CMB+BAO+SNIa(Dov) as a function of future cosmic time $t-t_0$ in units of $H_0^{-1}$, where $t_0$ is the present cosmic time. }
    \label{fig:ratio}
\end{figure}

The values of $\epsilon$ and $\delta$, in contrast, are well constrained within the positive range, with typical magnitudes of $\mathcal{O}(10^{-2})$ and $\mathcal{O}(10^{-1})$, respectively. Using DES-Dovekie leads to a mild shift toward larger values, in agreement with the enhancement of the new-physics signal discussed above. The positivity of $\epsilon$ and $\delta$ (together with $1+w_X<0$) ensures that the composite DE fluid behaves as quintessence at present and also the existence of the aforesaid crossing. In the middle plot of Fig. \ref{fig:tripleplot}, we show the shape of the individual contributions $\rho_X(z)$ and $\rho_{\rm vac}(z)$ to the total composite DE fluid $\rho_D(z)$, together with the latter, for each of the profiled values of $w_X$ (cf. again Fig. \ref{fig:profile_zcross}). Since $\nu<0$, the cosmon injects energy into the vacuum, making $\rho_{\rm vac}$ ($\rho_X$) to grow (decrease) with the cosmic expansion. The peak in $\rho_D(z)$ happens of course at $z^*$, when there is the crossing of the phantom divide and when the cosmon switches its character from quintessence to phantom matter -- its energy density decreases for all $z$ and goes from $\rho_X>0$ at $z>z^*$ to $\rho_X<0$ at $z<z^*$. \jtext{Despite that the evolution of the total DE density $\rD$ is slow and remains relatively flat, we can see from the middle panel of Fig. \ref{fig:tripleplot} that the evolution of the individual DE components is faster. Indeed,  as it nears $z=0$, the cosmon density plunges into the negative energy domain (transmuting into PM) and transferring its positive energy  to the vacuum, which raises fast towards our present. } The peak in $\rho_D$ is directly related to the one observed in the Hubble rate (see the bottom panel of Fig. \ref{fig:tripleplot}) and is similar to that found in the CPL model. Current uncalibrated BAO and SNIa data favor a departure from $\Lambda$CDM at $z\lesssim 1$, featuring a bump in $E(z)$. This bump is also observed in $H(z)$, of course, but its height depends on the (model-dependent) CMB calibration of the  cosmic ladders, which sets a preferred value of the sound horizon at the baryon-drag epoch $r_{\rm d}$ and $H_0$. Differences in these quantities (cf. Table \ref{tab:results}) lead to differences in the relative sizes of the bumps in $H(z)$ observed in the CPL and $\Lambda$XCDM models, while preserving the overall shape of $E(z)$. To keep the distance to the last scattering surface measured by Planck under the assumption of standard pre-recombination physics, this enhancement must be compensated at higher redshifts, making $H(z)$ to go below the $\Lambda$CDM curve. This is why the composite DE fluid has to behave as phantom DE at high $z$.

Let us comment now on BBN constraints from our posteriors. The condition $w_X-\epsilon<\frac{1}{3}$ is satisfied by all the points in the chains, so the function $R_N(z)$ from Eq. \eqref{eq:rBBN} is residual at $z\gg1$. The surviving term $r_\epsilon$ given by Eq.\,\eqref{eq:repsilon} is restricted to be $|r_\epsilon| \lesssim 0.01$. We find values of $r_\epsilon$ around this order of magnitude, with most of the points fulfilling the bound. More specifically, we find that its 1D marginalized posterior at $95\%$ CL is $r_\epsilon = -0.007\pm0.004$ with both DES-Dovekie and Pantheon+.

We also study what is the status of the cosmic coincidence problem in the light of the $\Lambda$XCDM model. This is a crucial point that cannot be addressed within the CPL framework, basically because it can not alleviate the problem. We perform this analysis by plotting the coincidence ratio given in Eq. \eqref{eq:rzMD} as a function of the future cosmic time  (cf. Fig. \ref{fig:ratio}). For all the profiled values of the cosmon EoS parameter $w_X<-1$ we find that $r(t)$ is bounded from above, but not only so. We find that for the most favored values of $w_X$ (i.e., those lying in the plateau of its profile likelihood), the aforementioned maximum is of order unity. Moreover, it is reached at a future time from now of the order of the current cosmic age. Therefore, we conclude that the $\Lambda$XCDM model is not only able to explain the crossing of the phantom divide, but also to alleviate the coincidence problem to a significant extent, without imposing any additional prior designed to force that.

The posterior values of $H_0$ displayed in Table \ref{tab:results} indicate that it is not possible to alleviate the Hubble tension within the $\Lambda$XCDM, at least with the data used here and under the assumption of $\wX=$const. If this condition were to be relaxed, the situation could change, but for the sake of the current work we remain with the simplest scenario, which yields values of the Hubble parameter below the $\Lambda$CDM and CPL, although still compatible with them at $1-2\sigma$ CL.

At very high
redshift (e.g. at CMB photon decoupling),  the total DE density \eqref{eq:ODz} asymptotes to $\OD(z)\simeq r_\epsilon (1+z)^{3(1+\wm)}$ in the relevant region, where $r_\epsilon$ was defined in Eq.\,\eqref{eq:repsilon}.
As a result, the Hubble rate \eqref{eq:FL2} takes on the asymptotic form
$H^2(z)\simeq H_0^2\,\hat{\Omega}_m^0\,(1+z)^{3(1+\wm)}$, in which
$\hat{\Omega}_m^0\equiv\Omo\,(1+r_\epsilon)$. The latter behaves as a sort of  ``renormalized''mass fraction parameter.
The two parameters $\hat{\Omega}_m^0$ and $\Omega_m^0$ coincide only for $\epsilon=0$.  Recall that $r_\epsilon$ is restricted by BBN bounds to satisfy $|r_\epsilon|\lesssim 1\%$, which we approximately saturate. Due to the fact that $\epsilon>0$ and $\wX<0$ in the relevant region of parameter space (cf. Table \ref{tab:results}),  we have $r_\epsilon<0$. Therefore, $\hat{\Omega}_m^0$ is reduced by about $1\%$ with respect to the $\Lambda$CDM value $\Omo$. Alternatively, one may keep the standard mass fraction and interpret this effect as a downward renormalization of the Hubble parameter: $H_0^2\to H_0^2 (1+r_\epsilon)<H_0^2$.

In more geometric terms, we can say that the lowering of $H_0$ is caused by an increase in the value of the ruler $r_{\rm d}$, which calibrates the ladders in the opposite direction to the one preferred by the local measurements from SH0ES \citep{Riess:2021jrx}. The sound horizon at decoupling increases because the Hubble expansion rate is reduced by two effects: (i) $r(z_{\rm dec})\sim -0.01$ (as noted above), implying that the composite DE density is negative there and amounts to approximately $1\%$ of the critical density. This is consistent with existing early DE bounds \citep{Gomez-Valent:2021cbe}; (ii) to produce the same amount of clustering during recombination, the non-relativistic matter density must be lower than in scenarios with a negligible DE fraction, such as $\Lambda$CDM or CPL, where $r(z_{\rm dec})\sim 10^{-9}$ (cf. Table \ref{tab:results}). The inability of late-time DDE models to alleviate the Hubble tension in the absence of new physics before recombination and in the presence of anisotropic BAO data has been demonstrated in numerous previous works (see, e.g., \cite{Sola:2017znb,Knox:2019rjx,Krishnan:2021dyb,Lee:2022cyh,Keeley:2022ojz,Gomez-Valent:2023uof,Gonzalez-Fuentes:2025lei,Gonzalez-Fuentes:2026rgu,Pedrotti:2025ccw,Bansal:2026axl,Sabogal:2026ipu}), although some room remains for very low-redshift solutions (see, e.g., \cite{Perivolaropoulos:2022khd,Alestas:2020zol,Alestas:2021luu,Gomez-Valent:2023uof}). Our results just point to the same direction.

Regarding the growth tension, the $\Lambda$XCDM model produces values of $\sigma_{12}\sim 0.81$ and $S_8\sim 0.825$ that are slightly larger than those predicted by the $\Lambda$CDM model. This is mainly because the current matter density fraction is larger and the dark energy density is more strongly suppressed at high redshifts due to the phantom behavior of the composite dark energy, as in the CPL.  In fact, the situation regarding these tensions is not different from the one encountered in the CPL, and an extended analysis is needed for a more detailed assessment of the status of the $H_0$ and growth tensions in the context of the $\Lambda$XCDM model. As explained in Sec. \ref{sec:DataAnalysis}, our main goal in this Letter is to show that there exists a wide region of the parameter space in the $\Lambda$XCDM model that is able to produce a crossing of the phantom divide and mitigate the coincidence problem, a  problem that affects not only the $\Lambda$CDM model, but also many DDE models, such as the CPL model.

Finally, it is worthwhile to emphasize that the achievements of the $\CC$XCDM demonstrated here have been obtained at a minimum cost of assumptions. In fact, we have left completely unspecified  the nature of the cosmon $X$ and we have just assumed that its EoS is constant ($\wX=$const.) and lies somewhere in the deep PM domain. With only this generic hypothesis, we obtain three successful accomplishments at  a time: i) crossing of the phantom divide by the effective EoS of the model ($\weff$) at the observed redshift range by DESI, ii) significant alleviation of cosmic coincidence; iii) a quality fit to the data which outperforms the standard $\CC$CDM as well as the CPL parameterization. Taking also into account that the core structure of the $\CC$XCDM is the running vacuum model -- based on QFT in curved spacetime\,\citep{SolaPeracaula:2022hpd,SolaPeracaula:2026pgi} -- and that the additional ingredient $X$ may represent a large family of existing generalized theories of gravity (see the Introduction), the number of obtained benefits is quite substantial and could possibly be further enhanced by specifying the nature of the cosmon and/or assuming some evolution of its EoS, $\wX=\wX(z)$, which might e.g. help resolving the main cosmological tensions. These extensions of the $\CC$XCDM capacities are left, of course, for future studies, which should also take into account a more complete set of cosmological data.


\section{Conclusions}\label{sec:Conclusions}

In this Letter, we have closely examined the currently hot issue associated with the  crossing of the phantom divide by the equation of state (EoS) parameter of dynamical dark energy (DDE), a conspicuous feature observed by the DESI Collab.\,\citep{DESI:2025zgx} with the help of phenomenological parameterizations, and reconfirmed by model-agnostic analyzes of the same data, see, e.g., \cite{Gonzalez-Fuentes:2025lei,Gonzalez-Fuentes:2026rgu}.  The main purpose of our work is to illustrate that such a crossing hallmark of the observed DDE can be accounted for by using a sound theoretical approach beyond simple parameterizations. Specifically, we have exploited the framework of the $\CC$XCDM,  a composite version of the Running Vacuum Model (RVM) \citep{SolaPeracaula:2022hpd,Sola:2013gha}. The RVM is based on fundamental principles such as quantum field theory (QFT) and string theory, and phenomenologically is known to help cure the cosmological tensions. It generally provides a good fit to the cosmological data\,\,\citep{SolaPeracaula:2023swx,SolaPeracaula:2021gxi} and the `flipped' variant of the RVM can even help describe the crossing feature\ \citep{deCruzPerez:2025dni}. However, the targeted approach that we have analyzed here, the $\CC$XCDM, performs optimally and at the same time can ease the cosmic coincidence problem.

In fact, the  $\CC$XCDM framework\,\citep{Grande:2006nn} was originally devised to address that problem. Although here we used it primary for a very different purpose,  as a bonus we find that in the very same region of the parameter space where the model can optimize the description of the crossing feature it can also greatly alleviate the cosmic coincidence problem. This double goal can be achieved thanks to the presence of an additional ingredient, the `cosmon' $X$, a generic entity that exchanges energy with the vacuum, and as a result vacuum and cosmon are both dynamical quantities. The cosmon, however, must possess a special attribute, which is to behave as `phantom matter' (PM), a cosmic fluid first introduced in\ \cite{Grande:2006nn}. Phantom matter shares the property $\wX<-1$ with ordinary phantom DE, but on the contrary it carries positive pressure and negative energy density: $\pX=\wX\rX>0$. This makes possible the existence of a turning point in the future evolution of the universe.

A remarkable feature of $\CC$XCDM is that in the domain of parameter space where there is crossing of the phantom divide, the effective EoS of the model, $\weff$, performs the crossing in only one direction, namely, from phantom DE in the past to quintessence-like behavior near our time (it is never the other way around). This EoS pattern is exactly what is observed by DESI. Furthermore, to achieve this goal it is not necessary to  specify the microscopic properties of the cosmon, except to assume that its EoS parameter must lie in the deep phantom domain, typically $\wX<-1.5$, as this is the region picked out by our fit (cf. Table \ref{tab:results}). The parameter space for the cosmon is therefore not restricted to a narrow stripe of that space, quite the opposite.  \jtext{Once $\wX$ is in that domain, there is no special preference for its value, as all of them entail a crossing point near the observed redshift range by DESI  -- cf. Figs.\,\ref{fig:profile_zcross} and \ref{fig:tripleplot} (upper plot).  This feature is also quite remarkable. Thus, no fine tuning of any kind is necessary to achieve the main aim}. Besides, \jtext{we wish to emphasize once more} that the nature of the cosmon also enjoys a large degree of freedom, as it can accommodate a wide number of possibilities. It  can be a  field, but in general it may stand for a term of the vacuum effective action of gravity that contributes  to the total value of the DE density of the universe beyond the usual vacuum energy associated with the cosmological term. The crossing of the phantom divide is then guaranteed provided $X$ behaves as PM and the vacuum itself evolves with the expansion along the lines of the RVM. Actually, there exist theoretical developments in the context of the stringy version of the RVM providing a raison d'\^etre for the notion of PM  at a fundamental level\ \citep{Mavromatos:2021urx,Mavromatos:2020kzj}, although in general such an unusual `demeanor' of the cosmon should not cause any theoretical uproar, given the aforesaid fact that the EoS parameter of $X$ does not necessarily correspond to a fundamental field. Several examples of gravity frameworks potentially providing effective cosmon-like candidates have been mentioned in the Introduction.

In our study, we have shown that $\CC$XCDM offers a rich scenario of possibilities for our understanding of the crossing feature of the DDE on more fundamental grounds and not just through efficient, though physically opaque, parameterizations of the DE. In the $\CC$XCDM, the RVM is the central theoretical structure, but a generic piece $X$ acting as phantom matter must be adjoined to ensure an interplay with the vacuum energy and trigger the correct crossing of the phantom divide.  In the stringy version of the RVM, the `bubbles' of PM (where the cosmon inhabits) are induced by quantum fluctuations associated with the nearing of the universe towards a de Sitter phase, which is just what our universe is doing now.  Since the pressure of the cosmon inside these bubbles is positive, this could trigger an anomalous outgrowth of structures at high redshifts, say in the range  $z\sim 5-10$, which might also help explain the appearance of the large scale structures recently spotted  by the JWST mission\,\citep{Labbe:2022ahb,Adil:2023ara,Menci:2024rbq}. Such overgrowth anomaly at large redshifts, which finds no explanation in the concordance $\CC$CDM, might also be described within the current proposal and could be induced by the quantum tunneling process towards the final de Sitter era \citep{Gomez-Valent:2024tdb}. Further studies should be devoted to this exciting possibility, which we mention here in passing as one more  potential spin-off benefit of our framework.

All in all, despite the traditional difficulties surrounding the notion of quantum vacuum, the RVM framework in combination with other, very much generic, elements of the effective action of vacuum borrowed from the large reservoir of modified theories of gravity existing today could eventually provide not only a possible solution to those fundamental conundrums, but also  to specific problems of modern cosmology, such as the current tensions and the observed phenomenological properties of the dynamical dark energy, what gives hope  for attaining a deeper understanding of the cosmological evolution of the universe from  fundamental principles.

\section{Acknowledgements}
This work is  partially supported by the projects PID2022-136224NB-C21, CEX2024-001451-M and by the grant 2021-SGR-249 (Generalitat de Catalunya). AGV is funded by “la Caixa” Foundation (ID 100010434)
and the European Union's Horizon 2020 
programme under the Marie Sklodowska-Curie grant agreement
(fellowship
code LCF/BQ/PI23/11970027). AGF is funded by the grant FPU24/01241 (MICIU). The authors acknowledge networking support by the COST Association Action CA21136 ``{Addressing observational tensions in cosmology
with systematics and fundamental physics (CosmoVerse)}'', and JSP also by CA23130 ”Bridging high and low energies in search of QG (BridgeQG)”.

\newpage

\appendix
\section{Full triangle plot and breakdown of $\chi^2_{\rm min}$ contributions}\label{sec:A}

\begin{figure}[ht!]
    \centering
    \includegraphics[width=\linewidth]{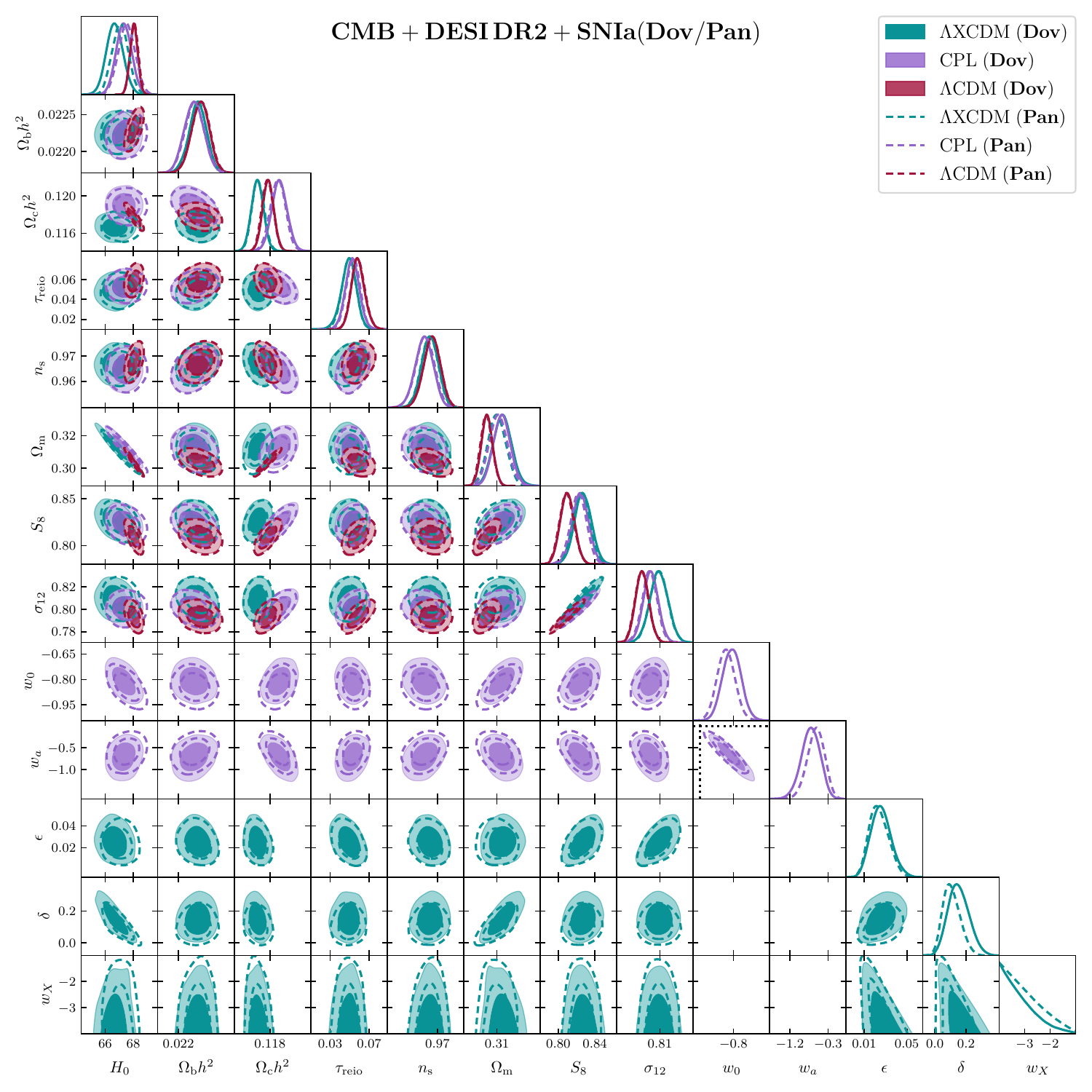}
    \caption{\scriptsize Full triangle plot for the various models studied in this paper. We show the constraints at 68$\%$ and $95\%$ CL in all the relevant planes of the parameter spaces, together with the individual one-dimensional posterior distributions. $H_0$ is given in km/s/Mpc.}
    \label{fig:full_triangle_plot}
\end{figure}

In Fig. \ref{fig:full_triangle_plot} of this appendix we provide a joint corner plot for the models studied, obtained with the 2 datasets described in Sec. \ref{sec:DataAnalysis}. In addition, in Table \ref{tab:table_chi2} we unfold all the individual likelihood contributions of the $\chi^2_{\rm min}$ of Table \ref{tab:results}.

\begin{table}[ht!]
   \centering
   \begin{tabular}{c||c|c|c||c|c|c}

    & \multicolumn{3}{c|}{\textbf{CMB + DESI DR2 + PantheonPlus}} & \multicolumn{3}{|c}{\textbf{CMB + DESI DR2 + DES Dovekie}} \\
   \hline
   $\chi^2_i$ & $\mathbf{\Lambda}$\textbf{CDM} & \textbf{CPL} & $\mathbf{\Lambda}$\textbf{XCDM} & $\mathbf{\Lambda}$\textbf{CDM} & \textbf{CPL} & $\mathbf{\Lambda}$\textbf{XCDM} \\ \hline \hline
   $\chi^2_\mathrm{SNIa}$ & $1406.28$ & $1403.08$ & $1402.91$ & $1634.43$ & $1629.33$ & $1628.93$ \\
   $\chi^2_\mathrm{BAO\, DESI\, DR2}$ & $11.48$ & $9.73$ & $9.38$ & $13.77$ & $9.56$ & $9.47$ \\
   $\chi^2_{\mathrm{Planck\, 2018\, low-\ell\, TT}}$ & $22.63$ & $22.81$ & $22.97$ & $22.78$ & $23.03$ & $22.95$ \\
   $\chi^2_{\mathrm{Planck\, 2018\, low-\ell\, EE}}$ & $396.52$& $396.81$ & $395.82$ & $396.08$ & $396.37$ & $396.82$\\
   $\chi^2_{\mathrm{Planck\, NPIPE\, high-\ell\, CamSpec \, TTTEEE}}$ & $10547.91$ & $10544.93$ & $10542.07$ & $10545.93$ & $10544.61$ & $10541.32$\\
   $\chi^2_\mathrm{Planck\, PR4\, lensing}$ & $9.07$ & $8.27$ & $8.72$ & $9.31$ & $8.30$ & $8.85$ \\ \hline
   $\chi^2_\mathrm{min}$ & $12393.89$ & $12385.64$ & $12381.87$ & $12622.30$ & $12611.22$ & $12608.36$ \\ \hline

\end{tabular}
\caption{\scriptsize Individual $\chi^2_i$ contributing to $\chi^2_{\rm min}$, obtained in the fitting analyzes  for the various models with CMB+BAO+SNIa.}
\label{tab:table_chi2}
\end{table}

\newpage

\section{BAO and SNIa distances}\label{sec:B}
\begin{figure}[ht!]
    \centering
    \includegraphics[width=0.75\linewidth]{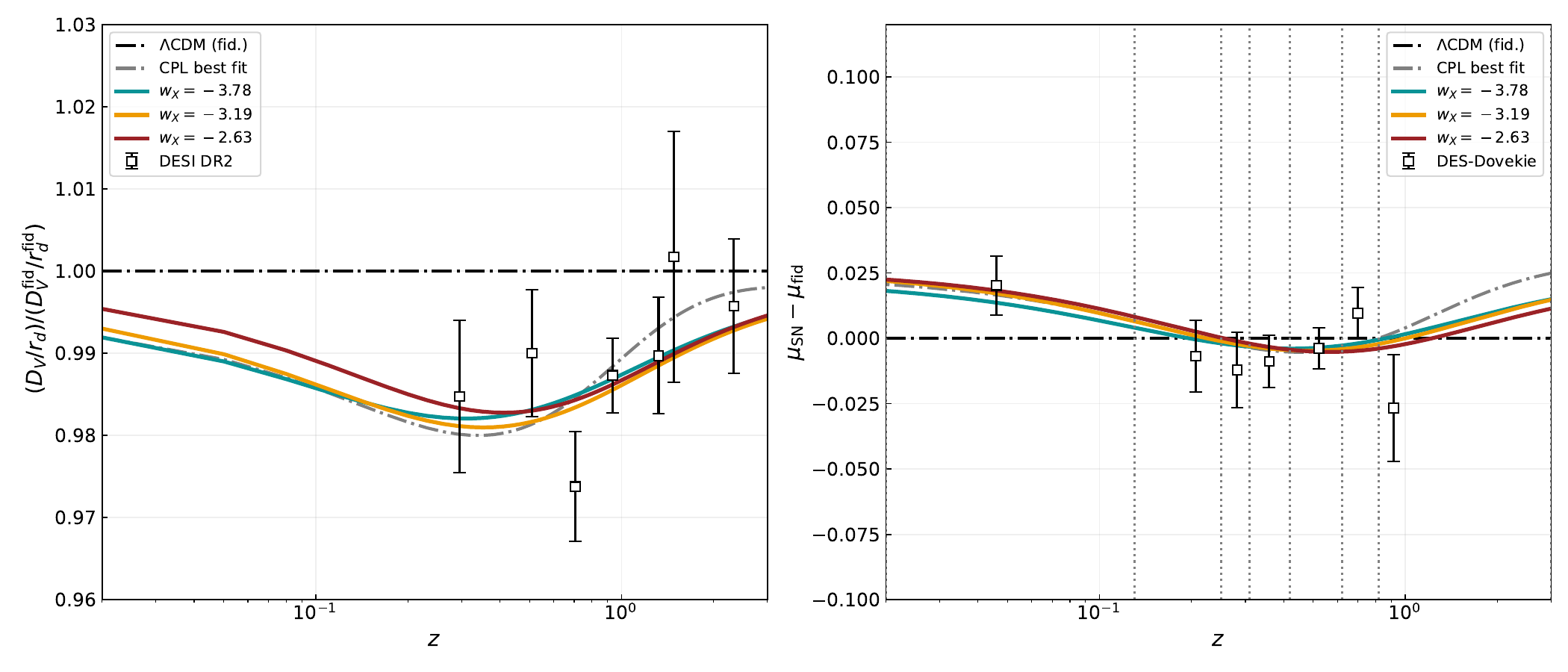}
    \caption{\scriptsize BAO and SNIa distances using DES-Dovekie. For SNIa distance modulus, we have binned the data as described in \citep{DESI:2025zgx}. We represent the curves obtained for different bins of the profile likelihood in $w_X$ to emphasize that different values of $w_X$ yield similar observables.}
    \label{fig:dist_dove}
\end{figure}

\begin{figure}[h!]
    \centering
    \includegraphics[width=0.75\linewidth]{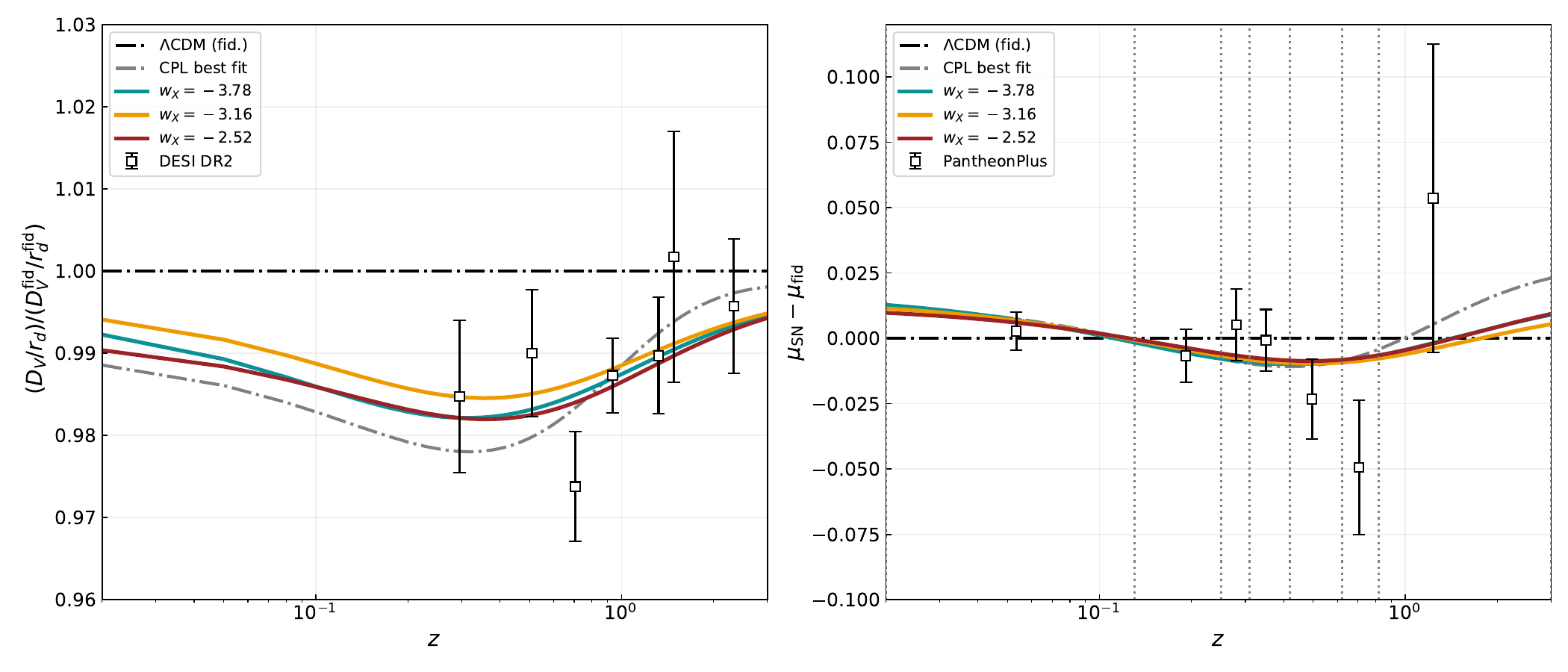}
    \caption{\scriptsize Same as Figure \ref{fig:dist_dove} but with PantheonPlus.}
    \label{fig:dist_PP}
\end{figure}

For completeness, in Figures \ref{fig:dist_dove} and \ref{fig:dist_PP} we display BAO angle-averaged distances $D_V/r_d$ from DESI DR2 and binned distance moduli $\mu$ from PantheonPlus/DES-Dovekie with respect to the fiducial {\it Planck} 2018 best fit $\Lambda$CDM. We compare the curves obtained for different profiled values of $w_X$. The similarity between the curves illustrates the point raised in Section \ref{sec:Discussion} about the insensitivity of $w_X$ in the range preferred by the data, due to shifts in $\nu$ and $\Omega_X^0$.

\bibliographystyle{aasjournal}

\end{document}